\documentclass[oneside, a4paper, onecolumn, 11pt]{article}
\usepackage[left=2.5cm,top=2.5cm,bottom=2.5cm,right=2.5cm]{geometry}
\usepackage{amsmath}
\usepackage{graphicx}
\usepackage{booktabs}
\usepackage{siunitx}
\usepackage{fancyhdr}
\usepackage{calc}
\usepackage{amssymb}
\usepackage{lipsum}
\usepackage{makecell}
\usepackage{titlesec}
\usepackage[super,sort&compress,comma]{natbib}
\usepackage{hyperref}
\hypersetup{
    colorlinks=true,
    linkcolor=black,
    filecolor=gray,
    urlcolor=blue,
    citecolor=black,
}
\usepackage{setspace}
\usepackage{amsfonts}
\usepackage{commath}
\usepackage[innercaption]{sidecap}
\usepackage[framemethod=TikZ]{mdframed}
\usepackage{hyperref}
\usepackage{parskip}
\usepackage{pifont}
\usepackage[labelfont=bf]{caption}
\usepackage{algorithm}
\usepackage{algorithmicx}
\usepackage{algpseudocode}

\newcommand{\Rom}[1]{\expandafter\@slowromancap\romannumeral #1@}
\makeatother
{ \normalfont} 
\DeclareMathOperator*{\argmin}{argmin}

\expandafter\def\expandafter\normalsize\expandafter{%
	\normalsize
	\setlength\abovedisplayskip{0pt}
	\setlength\belowdisplayskip{5pt}
	\setlength\abovedisplayshortskip{0pt}
	\setlength\belowdisplayshortskip{5pt}
}

\definecolor{Gray}{gray}{0.75}
\definecolor{changecolour}{rgb}{0, 0, 0.8}

\newmdenv[backgroundcolor=Gray, leftmargin = 0pt, rightmargin = 0pt, linewidth = 0pt, roundcorner = 2 pt, innerleftmargin=5pt, innerrightmargin=5pt, innertopmargin=5pt, innerbottommargin=5pt]{Frame}

\begin{document}

\newcommand{\kk}{\langle k \rangle}
\newcommand{\kkk}{\langle k^2 \rangle}
\newcommand{\er}{Erd\H{o}s-R\'{e}nyi}
\newcommand{\red}{\color{red}\footnotesize}
\newcommand{\blue}[1]{{\color{blue} #1}}
\newcommand{\subfigimg}[3][,]{%

	\setbox1=\hbox{\includegraphics[#1]{#3}}
	\leavevmode\rlap{\usebox1}
	\rlap{\hspace*{30pt}\raisebox{\dimexpr\ht1-2\baselineskip}{#2}}
	\phantom{\usebox1}
}

\linespread{1.2}

\begin{center}
{\LARGE \textbf{Autonomous inference of complex network dynamics from incomplete and noisy data}}

\vspace{2mm}
Ting-Ting Gao$^{1,2}$ and Gang Yan$^{1,2,3,*}$
\end{center}

\small{
\begin{enumerate}
\item
\textit{MOE Key Laboratory of Advanced Micro-Structured Materials and School of Physics Science and Engineering, Tongji University, Shanghai 200092, P. R. China}
\item
\textit{Frontiers Science Center for Intelligent Autonomous Systems, Tongji University, Shanghai 200092, P. R. China}
\item
\textit{Center for Excellence in Brain Science and Intelligence Technology, Chinese Academy of Sciences, Shanghai 200031, P. R. China}
\end{enumerate}
\begin{itemize}
\item[*]
Correspondence to Gang Yan (gyan@tongji.edu.cn).
\end{itemize}
}

\vspace{5mm}

\noindent
\textbf{The availability of empirical data that capture the structure and behavior of complex networked systems has been greatly increased in recent years, however a versatile computational toolbox for unveiling a complex system's nodal and interaction dynamics from data remains elusive. Here we develop a two-phase approach for autonomous inference of complex network dynamics, and its effectiveness is demonstrated by the tests of inferring neuronal, genetic, social, and coupled oscillators dynamics on various synthetic and real networks. Importantly, the approach is robust to incompleteness and noises, including low resolution, observational and dynamical noises, missing and spurious links, and dynamical heterogeneity. We apply the two-phase approach to inferring the early spreading dynamics of H1N1 flu upon the worldwide airline network, and the inferred dynamical equation can also capture the spread of SARS and COVID-19 diseases. These findings together offer an avenue to discover the hidden microscopic mechanisms of a broad array of real networked systems.}

\vspace{5mm}

\section*{Introduction}
From two-photon calcium imaging of neuronal activities\cite{grewe2010high,stetter2012model}, high-throughput genetic experiments\cite{reuter2015high,levy2016advancements} to digital recordings of human mobility\cite{colizza2006role,brockmann2013hidden,chang2021mobility}, our ability to observe the dynamic behavior of nodes in complex biological, social and technological systems has advanced spectacularly in the past years. The collected observations, often in the form of time-series data, allow us to extract the dynamic patterns of a system's individual nodes. To gain meaningful insights on the system, however, such a reductionist approach of tracking all individual nodes is insufficient. Indeed, complex system behavior emerges not just from the single nodes, but rather from the dynamic interactions between the nodes\cite{newman2011structure,barzel2013universality,harush2017dynamic,stankovski2017coupling,breakspear2017dynamic,santolini2018predicting,buldyrev2010catastrophic,yang2017small,pastor2015epidemic,brockmann2013hidden,castellano2009statistical,becker2017network}. This requires us to infer complex network dynamics, \textit{i.e.} to retrieve both self nodal dynamics and interaction dynamics from the accumulating data of network topological structure and nodes' activities.

The balance of self vs.\ interaction dynamics is most naturally captured by a general equation that tracks the activities of all nodes via\cite{barzel2013universality}
\begin{equation}
   \frac{\text{d}{\textbf{x}_i}(t)}{\text{d}{t}}=\textbf{F}\big(\textbf{x}_i(t)\big)+\sum_{j=1}^{n}A_{ij}\textbf{G}\big{(}\textbf{x}_i(t),\textbf{x}_j(t)\big{)},
\label{EquDyn}
\end{equation}
where $\textbf{x}_i(t) \equiv (x_{i,1}(t), \ldots, x_{i,d}(t))^{\top}$ is node $i$'s $d$-dimensional activity, representing, \textit{e.g.}\ the membrane potential of a neuron in a brain network\cite{barzel2013universality,breakspear2017dynamic}, the proportion of infected people in a country or region\cite{colizza2006role,brockmann2013hidden,chang2021mobility} or the state of a component in an oscillator network\cite{arenas2008synchronization}. These activities are driven by the self-regulation function $\textbf{F}(\textbf{x}_i) \equiv  (F_1(\mathbf{x}_i),\ldots,F_d(\mathbf{x}_i))^{\top}$, designed to describe the dynamics of all nodes in isolation, and by the pairwise function $\textbf{G}(\textbf{x}_i(t),\textbf{x}_j(t)) \equiv (G_1(\mathbf{x}_i,\mathbf{x}_j),\ldots,G_d(\mathbf{x}_i,\mathbf{x}_j))^{\top}$ which captures the dynamic mechanisms of interaction between the nodes. Finally, the network $A_{ij}$, an $n \times n$ adjacency matrix, denotes the influence or flow from node $j$ to $i$, where $n$ is the number of nodes in the system. As shown by Barzel and Barab\'{a}si, with appropriate choices of the nonlinear functions $\textbf{F}$ and $\textbf{G}$, Eq.\ (\ref{EquDyn}) is able to describe a broad range of complex systems\cite{barzel2013universality}. However, for most real systems, the functions $\textbf{F}$ and $\textbf{G}$ are unknown. Hence, a pressing lacuna in the study of complex systems is a versatile computational toolbox for automatically inferring Eq.\ (\ref{EquDyn}) from the observed data of network topology $A_{ij}$ and node activities $\textbf{x}_i(t)$.

Complex biological, social or technological systems lack the fundamental physical rules that govern particle systems, so we do not have a \textit{priori} knowledge of their internal microscopic mechanisms\cite{barzel2015constructing}. Therefore, the goal is not to only identify the model's parameters but rather to retrieve the forms of $\textbf{F}$ and $\textbf{G}$ and infer the explicit model itself.
Despite the recent significant progress in developing methods to infer the governing equations of single- or few-body dynamics\cite{schmidt2009distilling,wang2011predicting,brunton2016discovering,rudy2017data,udrescu2020ai,raissi2018hidden,iten2020discovering}, the task of inferring network dynamics poses particular challenges. For example, $\textbf{F}$ and $\textbf{G}$ are usually of different types hence one cannot obtain their compact forms if using only orthogonal basis functions\cite{wang2011predicting,brunton2016discovering,frishman2020learning,bruckner2020inferring}; Node activities data are noisy and the mappings of network topologies are usually incomplete\cite{shandilya2011inferring,newman2018network}; Collective behavior, such as synchronization and consensus\cite{arenas2008synchronization}, can conceal the specific forms of microscopic mechanisms in interaction dynamics. To overcome these challenges we propose here a two-phase inference approach. Our analysis indicates that the two-phase strategy allows us to achieve efficient and, most importantly, highly accurate inference, even in the face of unfavorable scenarios, such as noisy or low-resolution data or an only partially mapped topology (Fig.\ 1a).



\vspace{5mm}

\section*{Results}
\subsection*{Overview of the two-phase inference approach}

Lacking a \textit{priori} knowledge of the structure of $\textbf{F}$ and $\textbf{G}$, a natural approach is to pre-construct two extensive libraries $L_F$ and $L_G$ that contain a variety of elementary functions. The combinations of these elementary functions can potentially generate the true network dynamics. In this work, the libraries contain not only orthogonal basis functions but include polynomials, trigonometric, exponential, fractional, rescaling, sigmoid and other activation functions that frequently used in various domains (see Supplementary Tables 1 and 2). Large libraries are helpful for finding a compact and optimal model to capture network dynamics but also make the inference problem more difficult, because, due to the lack of orthogonality, the elementary functions can be similar with each other and thus less discriminative.

Introducing the time series data $\textbf{x}_i(t)$, where $i=1,2,\ldots,n$, into $L_F$ and $L_G$, we obtain two time-varying matrices $\Theta_F(t)\equiv L_F(\textbf{x}_i(t))$ and $\Theta_G(t)\equiv L_G(\textbf{x}_i(t),\textbf{x}_j(t))$ that encode the patterns of node activities imposed by the elementary functions in $L_F$ and $L_G$ (Fig.\ 1b). Then the inference problem can be recast to selecting appropriate patterns in $\Theta_F(t)$ and $\Theta_G(t)$ that best match the evolution of observed system state $\Dot{\textbf{x}}(t)$, \textit{i.e.} to inferring the sparse coefficients $\boldsymbol{\xi}_F$ and $\boldsymbol{\xi}_G$ that best solve
\begin{equation}
   \Dot{\textbf{x}}(t)=\widetilde{\Theta}_F(t)\boldsymbol{\xi}_F+\widetilde{A}\widetilde{\Theta}_G(t)\boldsymbol{\xi}_G,
\label{Form}
\end{equation}
where $\widetilde{A}\equiv A\otimes I_d$, $\widetilde{\Theta}_F \equiv \Theta_F\otimes I_d$, $\widetilde{\Theta}_G \equiv \Theta_G\otimes I_d$, symbol $\otimes$ denotes Kronecker product and $I_d$ is the $d$-dimensional identity matrix.
Here, we consider the general setting where each node state is $d$-dimensional and the network is directed and heterogeneous. Consequently, the problem of inferring complex network dynamics is high-dimensional and irreducible. Indeed, the number of elementary functions in $L_F$ and $L_G$ is approximately $25$, $80$ or $140$ when node activity itself has one, two, or three dimensions respectively in the simulation validations below (see Supplementary Tables 1 and 2).

Our approach is a two-phase procedure consisting of global regression and local fine-tuning. In phase I, we approximate the derivatives $\Dot{\textbf{x}}(t)$ (see Methods) and calculate the matrices $\widetilde{\Theta}_F(t)$ and $\widetilde{\Theta}_G(t)$, and then normalize each column of them (Fig. 1b). These normalized data are used to identify, through regression, the leading elementary functions that are most probably constituents of the true $\textbf{F}$ and $\textbf{G}$ (Fig.\ 1c, and see also Methods). Phase I is able to significantly narrow down the model space but the dynamical equation inferred by such regression alone lacks generative power (Fig. 1d). Next, in phase II, we perform fine-tuning with the original values of $\Dot{\textbf{x}}(t)$, $\widetilde{\Theta}_F(t)$ and $\widetilde{\Theta}_G(t)$, \textit{i.e.}\ without normalization. We use topological samplings (see Methods) and the weighted Akaike's information criterion (wAIC, see Methods) to sequentially remove the elementary functions with smallest inferred coefficients (Fig. 1e). The final sets of elementary functions and their coefficients $\hat{\boldsymbol{\xi}}_F$ and $\hat{\boldsymbol{\xi}}_G$ compose $\hat{\textbf{F}}$ and $\hat{\textbf{G}}$, leading to the inferred dynamics of complex networks (Fig. 1f).

\subsection*{Inferring complex network dynamics}

To validate the effectiveness of our approach we apply it to infer five network dynamics, including Hindmarsh-Rose\cite{rabinovich2006dynamical} (HR, $d=3$) and FitzHugh-Nagumo\cite{rabinovich2006dynamical} (FHN, $d=2$) neuronal systems, social balance dynamics\cite{marvel2011continuous} (SB, $d=1$), Kuramoto dynamics\cite{strogatz2001exploring} ($d=1$), and coupled heterogeneous R\"ossler oscillators\cite{barahona2002synchronization} ($d=3$), here $d$ is the dimension of each node activity. To obtain the nodes' activities data we respectively simulate these dynamics (Supplementary Table 4) on a variety of toplogies, including Erd\H{o}s-R\'enyi (ER) and scale-free (SF) synthetic networks, and five empirical networks --  cellular-level brain networks of {\it C. elegans} and {\it Drosophila}, Advogato social network, power grids of North-Europe and United States. The time series of node activities and each network topology are the input data to our approach. The five specific equations governing these dynamics are the ground truths that we aim to infer. These dynamical models and networks are widely used in various domains and exhibit different properties (see Supplementary Information Sections II and III), accounted for the diversity of our tests.

Figure\ 2 illustrates the procedure of inferring FitzHugh-Nagumo neuronal network dynamics. Through the global regression, Phase I identifies ten most relevant elementary functions for each dimension of FHN (Fig.\ 2b), and then, by local fine-tuning, Phase II autonomously learns the compact and optimal form of the dynamical equation as well as the most appropriate coefficient for each of the necessary elementary functions (Fig.\ 2c). The form of the inferred equation in Fig.\ 2c perfectly matches the ground truth in Fig.\ 2a and the learnt coefficients are also highly accurate. Indeed, the relative errors $\Delta=(\xi-\hat{\xi})/\xi$, where $\xi$ and $\hat{\xi}$ are true and learnt coefficients respectively, are smaller than $3\%$ (Fig.\ 2d). The dynamical equation inferred by our approach exhibits generative power, being able to generate nodes' activities and trajectories that agree well with observation data (Fig.\ 2(e,f)).

Our approach also successfully infers the equations governing other four network dynamics. Regarding the accuracy of learnt coefficients, the relative errors $\left|\Delta\right| < 3\%$ for the Hindmarsh-Rose dynamics (Fig.\ 3a) and the edge dynamics (Fig.\ 3c) on both synthetic and empirical networks. In Kuramoto dynamics and coupled heterogeneous R\"ossler oscillators the self-dynamics are nonidentical, \textit{i.e.}\ each node's dynamics has its own form (see Supplementary Information Section\ III). Hence we aim to infer an \textit{effective} form of Eq.\ (1) that minimizes the inconsistency between inferred and true nodes' activities. Even for these more challenging cases, the two-phase approach still succeeds with relative coefficient errors $\left|\Delta\right| < 5\%$ or $<20\%$ (Fig.\ 3(e,g)). Both activities and trajectories generated by the effective equations exhibit high agreement with the true averaging dynamics (Fig.\ 3(f,h,i)).

\subsection*{Inferrability of network dynamics}

Whether a network dynamics is inferrable depends on several factors. Here we explore three key factors, namely synchronized dynamics, dynamical heterogeneity, and deficient libraries.

Synchronized dynamics. If a network is completely synchronized, \textit{i.e.}\ all of its nodes behave in the same manner\cite{strogatz2001exploring,barahona2002synchronization,arenas2008synchronization}, distinguishing the activities of a node and its neighbors becomes impossible and the microscopic interacting mechanism $\textbf{G}(\textbf{x}_i,\textbf{x}_j)$ between nodes will be cloaked and undiscoverable. In other words, more synchronized is a network, more difficult to infer its dynamics. Here we tune the coupling strength between nodes to change the degree of network synchronization (\textit{i.e.}\ the order parameter $\langle R\rangle$, see Supplementary Information Section IV), and test the capability of our two-phase approach in inferring partially synchronized network dynamics. As shown in Fig.\ 4a, although the inference inaccuracy increases when the system becomes more synchronized, our approach still can infer the true FHN equation even when the network is highly synchronized ($\langle R\rangle \approx$ 0.7). The inference inaccuracy is quantified by symmetric mean absolute percentage error (sMAPE, see Methods). The more accurate the inference result, the closer the sMAPE value to zero.

Dynamical heterogeneity. Equation (1) assumes that nodes have the same form $\textbf{F}$ of self-dynamics, yet this is not always true. For instance, although the self-dynamics of Kuramoto model is simply one elementary function $\omega$ representing the natural frequency of a node, different nodes can have different values of $\omega$. For such nonidentical self-dynamics it is difficult, if not impossible, to infer a specific form $\textbf{F}_i(\textbf{x}_i)$ for each node $i$ due to an $n$-fold increase in the dimensionality of potential model space ($n$ is network size). Therefore we aim to infer an \textit{effective} equation that best captures the {\it averaging} dynamics, as shown in Fig.\ 3(e,g). Here we further explore to what extent of dynamical heterogeneity our approach can tolerate. To do so we assign each node a value of $\omega$ randomly drawn from a normal distribution $\mathcal{N}(0,\sigma)$ and increase the standard deviation $\sigma$. The inference inaccuracy indeed increases when $\sigma$ becomes larger, and the two-phase approach can tolerate dynamical heterogeneity $\sigma \leq 0.5$ (Fig.\ 4b).

Deficient libraries. Although two rather comprehensive libraries of elementary functions are built, it is still possible that some elementary functions of the true unknown dynamics are missing. Another possibility is that the compact form of true dynamics cannot be composed by these elementary functions. For these cases, our two-phase approach will infer an alternative equation to capture the system behaviors. We test such capability in gene regulation and Hindmarsh-Rose neuronal dynamics whose true coupling functions are intentionally removed from $L_G$. As shown in Fig.\ 4c, the trajectories generated by the inferred and the true equations are close to each other, and the discrepancy is small for all nodes (see Methods and also Supplementary Information Section IV-B).

\subsection*{Inferring from incomplete and noisy data}

Incompleteness of mapped network topology and noises of observed nodes' activities are inevitable in real data\cite{shandilya2011inferring,newman2018network}. Hence, here we validate the robustness of our two-phase approach against low resolution, dynamical and observational noises, spurious and missing links, as well as through comparisons with previous methods\cite{brunton2016discovering,mangan2017model,casadiego2017model}.

Low resolution. Experimental and digitally recording technologies often have limited measurement frequencies, inducing low resolution of observed time series. To validate our approach's robustness against low resolution we numerically simulate the five nonlinear network dynamics in Figs.\ 2 and 3 with step size $0.01$, and then regularly down-sample the activities data. We calculate the failure ratios in inferring the form of true equations (Supplementary Figure 14a) and also the inference inaccuracies (Fig.\ 5a). The results show that the two-phase strategy requires only a proportion of $5\%$ to $50\%$ data for the inference.

Observational and dynamical noises. Observational noises are induced by the measuring process and dynamical noises represent the intrinsic stochasticity in dynamics. To produce the former we add Gaussian noises to nodes' activity data and quantify the intensity of observational noise with signal-to-noise-ratio (SNR, see Supplementary Information Section V-A); To imitate the latter we add a stochastic term of Gaussian white noise with intensity $\eta$ into the true dynamical equations and generate the nodes' activities data by numerical simulations of these stochastic differential equations (see Supplementary Information Section V-A). We test the impact of these two types of noises on the performance of the two-phase inference approach, without any denoising preprocess. As shown in Fig.\ 5b and Supplementary Figure 14b, the approach can tolerate dynamical noise with $\eta\leq 0.15$, meaning that it successfully reconstructs the hidden equations when the stochastic intensity is not higher than 15\% of the average amplitude of true deterministic dynamics. Moreover, the approach can tolerate $30$-dB observational noise (Fig.\ 5c and Supplementary Figure 14c).

Spurious and missing links. Spurious and missing links in real data induce an incomplete network topology $A_{ij}$, which further lead to an inaccurate interaction matrix $\Theta_G$. To test the impact of these erroneous links we randomly add or remove a fraction of links from the true network topology that was used to simulate the nodes' activities. Owing to the topological sampling in Phase II, our approach is able to tolerate $25\%$ spurious and $30\%$ missing links (Fig.\ 5(d,e) and Supplementary Figure 14(d,e)).

Comparison with previous methods. Two most illuminating and effective methods for dynamics inference are SINDy\cite{brunton2016discovering} and ARNI\cite{casadiego2017model}. Note that ARNI originally aimed at inferring network topology but can be transferred to infer network dynamics by minor modification (see Supplementary Information Section \uppercase\expandafter{\romannumeral5}-C). Here we compare our approach with SINDy and ARNI from different aspects, including the amount of required data (Fig.\ 5f), the robustness against observational noise (Fig.\ 5g), correlated dynamical noise (Fig.\ 5h and also Supplementary Information Section V-A), missing links (Fig.\ 5i), and different network sizes (Fig.\ 5j). While ARNI needs fewer data points if network topologies are complete and nodal activities do not have any noise (Fig.\ 5f), the two-phase approach outperforms both SINDy and ARNI in inferring complex network dynamics from incomplete and noisy data (Fig.\ 5g-j). We also perform comparisons with SINDy's variant\cite{mangan2017model} regarding partially synchronized or heterogeneous dynamics (Supplementary Figures 13 and 17). These results indicate that our approach can better handle high-dimensional networked systems and better cope with incompleteness and noises in data.

Ablation studies. Besides the two-phase strategy, our approach also involves three important components, namely, normalization in the first phase yet non-normalization in the second for solving the issue raised by highly skewed observations at different nodes, topological sampling for imitating the feature of observed incomplete topologies, and optimal selection by wAIC for determining the most appropriate complexity of inferred dynamics. The essentiality of the two-phase strategy and the three components is demonstrated by ablation studies. Specifically, we ablate each phase or component and then assess the performance of degenerated approaches. As shown in Fig.\ 5(k,l) and Supplementary Information Section V-B, the inference inaccuracy sMAPE indeed significantly increases if the phases or components are individually ablated.

\subsection*{Inference of empirical systems}

To demonstrate the approach's ability of handling empirical systems, we apply it to infer the spreading dynamics of infectious disease H1N1. The network underlying this diffusion system is the worldwide airline network, which captures the human mobility between different countries or regions and plays a dominant role for global disease spreading\cite{colizza2006role,brockmann2013hidden}. Each entry $A_{ij}$ of the weighted network's adjacency matrix $A$ represents the traffic volume from node $j$ to $i$, where each node denotes a country or region. The total passengers daily are approximately $\Phi=8.9\times10^6$ and, taking into account the population $P_i$ of each node $i$, the adjacency matrix is modified to
\begin{equation}
    \hat{A}_{ij}=\frac{\Phi}{\sum_{i=1}^nP_i}A_{ij}.
\end{equation}
The magnitude order of entries in matrix $\hat{A}$ is around $10^{-2}$ to $10^{-3}$. The nodal activities $x_i(t)$ are extracted from the daily reports of infected cases in each country or region. Here we consider the nodes whose accumulated H1N1 cases are more than $100$ and focus on the early spreading dynamics, \textit{i.e.}\ within the $45$ days since the first case was reported in each node, which captures the system behaviors before government control.

Based on these empirical data, our approach successfully infers a concise \textit{effective} dynamical equation
\begin{equation}
    \frac{\text{d}x_i}{\text{d}t}=ax_i+b\sum_{j=1}^N\hat{A}_{ij}\frac{1}{1+e^{-(x_j-x_i)}},
    \label{earlyspread}
\end{equation}
where $a=0.074$ and $b=7.130$ (see Supplementary Information Section \uppercase\expandafter{\romannumeral6} and Supplementary Figure 18). It is interesting that our approach infers a sigmoid (nonlinear) form, rather than the linear form of epidemic models, to better capture the interaction dynamics. This might be caused by the fact that people usually consciously travel less if their countries/regions or the destinations have high infection risk. While Eq.\ \ref{earlyspread} describes the dynamics of all nodes with the same parameters $a$ and $b$, we also extend it by taking into account of dynamical heterogeneity in the nodes, \textit{i.e.}\ to obtain $a_i$ and $b_i$ from each node $i$' activity data (see Fig.\ 6(b-e) and also Supplementary Figure 18).

Because empirical systems lack ground-truths, we verify the inferred Eq.\ \ref{earlyspread} by testing its generalizability to SARS and COVID-19 diseases. Based on the daily reported numbers within the first $45$ days in each node, we find that Eq.\ \ref{earlyspread} is also able to capture the early spread of SARS and COVID-19 upon the worldwide airline network. Indeed, as shown in Fig.\ 6(f-i) and Supplementary Figures 19-20, the evolution of cumulative numbers of SARS cases (for nodes whose eventual infected cases are more than 100)  and COVID-19 cases (for nodes whose eventual infected cases are more than 2000) agree well with the activities generated by Eq.\ \ref{earlyspread} with heterogeneous parameters $a_i$ and $b_i$.

\section*{Discussion}

Many real networks have been mapped so far but there are still complex systems whose network structure information is totally missing. For the later cases, a possible scheme is inferring their topological structure, especially directed or causal networks\cite{runge2019detecting,sugihara2012detecting, sun2015causal,kralemann2014reconstructing,shandilya2011inferring,casadiego2017model}, from nodes' activities data first and then applying our approach to infer system dynamics. It is worth noting that inferring network structure from nodes' activities data is also challenging, especially when the number of nodes is large\cite{frassle2017regression, gilson2016estimation}, because the number of parameters needing to be estimated is about $n^2$ where $n$ is network size. Therefore, how to simultaneously infer both structure and dynamics of large complex systems is still an outstanding problem.

Our work also raises several questions worthy of future pursuit. First, stochasticity in the dynamics of some real complex systems might be stronger than that we considered in this work. Such highly stochastic systems are better described by stochastic differential equations\cite{deco2009stochastic,bruckner2020inferring,genkin2021learning,zhao2021inferring}. Second, our approach does not account for discrete or boolean dynamics, or systems that contain thresholding terms or exhibit irregular dynamics with instability properties\cite{jahnke2008stable}. Third, when nodal activity is multi-dimensional, experimental access might be limited to a sub-dimension of the activity vector. The Koopman operator and time-delay embedding techniques are helpful for capturing the dynamical properties of sub-dimension observable systems\cite{champion2019discovery}. Yet, the problem remains unsolved for complex networked systems. Finally, the nodes in a complex system can have higher-order, beyond pairwise, couplings and such higher-order interactions may significantly impact the dynamics of networked systems\cite{battiston2021physics,lambiotte2019networks}. Hence it is an interesting direction to extend the approach to inferring higher-order network dynamics.

\vspace{8mm}
\section*{Methods}
\subsection*{Two-phase inference approach}
The left side of Eq.\ (\ref{EquDyn}) represents the time-varying derivative of each node's activity, which can be numerically obtained from $\textbf{x}_i(t)$ through the five-point approximation\cite{sauer2012numerical}
\begin{equation}
  \Dot{x}_t\approx\frac{x_{t-2\delta t}-8x_{t-\delta t}+8x_{t+\delta t}-x_{t+2\delta t}}{12\delta t},
\end{equation}
where $\delta t$ is the time step. Hence the specific goal is to infer both the exact structure and the corresponding coefficients of the self-dynamics function $\textbf{F}(\textbf{x}_i(t))$ and the interaction dynamics function $\textbf{G}(\textbf{x}_i(t),\textbf{x}_j(t))$.

Because we lack a {\it priori} knowledge of the forms of $\textbf{F}$ and $\textbf{G}$, hence we construct two comprehensive libraries, $L_F$ and $L_G$, for self- and interaction dynamics respectively, including polynomial, trigonometric, exponential, fractional, rescaling, and various activation functions as listed in Supplementary Tables 1 and 2. Introducing the observed time series of nodes' activities to the elementary functions in $L_F$ and $L_G$, we obtain two matrices $\Theta_F(t)=L_F(\textbf{x}_i(t))$ and $\Theta_G(t)=L_G(\textbf{x}_i(t),\textbf{x}_j(t))$ that describe the corresponding behaviors of these elementary functions (Supplementary Figure 1). To infer the compact forms that best match Eq.\ (\ref{Form}) we propose a two-phase approach.

Phase \uppercase\expandafter{\romannumeral1}, global regression. The purpose of this phase is to assess the relevance of each elementary function in $L_F$ and $L_G$ to the true, yet unknown, network dynamics. Given the observations of $\textbf{x}_i(t)$ for all $i$ at time $t$ we approximate the derivatives $\Dot{\textbf{x}}(t)$ and calculate the matrices $\widetilde{\Theta}_F(t)$ and $\widetilde{\Theta}_G(t)$. These values are highly skewed and can span several orders of magnitude (Supplementary Figure 3) due to the skewness of node degrees and the nonlinearity of system dynamics, which could induce overestimation of the importance for inherently low-value constituents. To eliminate this severe effect it is crucial to normalize each column in $\Dot{\textbf{x}}(t)$, $\widetilde{\Theta}_F(t)$ and $\widetilde{\Theta}_G(t)$. Then, the inference problem described by Eq.\ (\ref{Form}) is further recast to an optimization formula
\begin{equation}
   \argmin_{\boldsymbol{\xi}_F, \boldsymbol{\xi}_G}\int_0^T\big(\|{\widetilde{\Theta}_F(t)\boldsymbol{\xi}_F+\widetilde{A}\widetilde{\Theta}_G(t)\boldsymbol{\xi}_G - \Dot{\textbf{x}}(t) }\|^2\big)\text{d}t + \lambda (\left\|{\boldsymbol{\xi}_F}\right\| + \left\|{\boldsymbol{\xi}_G }\right\|),
\label{Objective}
\end{equation}
where $\lambda>0$ is a hyper-parameter that regulates the sparsity of coefficient vectors $\boldsymbol{\xi}_F$ and $\boldsymbol{\xi}_G$. We employ the regression analysis method of least absolute shrinkage and selection operator (\textit{i.e}.\ lasso) to solve (\ref{Objective}) and perform 5-fold validation to obtain the most appropriate value of $\lambda$ (see Supplementary Information Section I-B Algorithm 1). The resultant $\boldsymbol{\xi}_F$ and $\boldsymbol{\xi}_G$ capture the relevance of each elementary function in $L_F$ and $L_G$, enabling the identification of leading elementary functions that are most probably constituents of the true $\textbf{F}$ and $\textbf{G}$ (Fig.\ 1c). Consequently, Phase \uppercase\expandafter{\romannumeral1} is able to significantly narrow down the model space. However, the dynamical equation inferred by such regression alone lacks generative power. For instance, as shown in Fig.\ 1d, the trajectory generated by an inferred dynamical equation of Phase \uppercase\expandafter{\romannumeral1} deviates from that of the true network dynamics.

Phase \uppercase\expandafter{\romannumeral2}, local fine-tuning. In order to reconstruct generative and concise expressions for $\textbf{F}$ and $\textbf{G}$ we next perform fine-tunning in the reduced model space (see Supplementary Information Section I-B Algorithm 2). In contrast to Phase \uppercase\expandafter{\romannumeral1} we now use the original values of $\Dot{\textbf{x}}(t)$, $\widetilde{\Theta}_F(t)$ and $\widetilde{\Theta}_G(t)$, \textit{i.e.}\ without normalization, to further identify the necessary elementary functions and learn their precise coefficients. Since spurious or missing links in the observed network topology have an adverse effect on the learning, we perform topological sampling (see Methods) that imitates the feature of observed, usually incomplete, topologies. Another issue is to determine the minimal number of elementary functions required for reconstructing $\textbf{F}$ and $\textbf{G}$. To do so, we sequentially remove the elementary functions with smallest inferred coefficients and calculate, using a weighted version of Akaike's information criterion (wAIC, see Methods), the information inconsistency between the observed nodes' activities and the remaining set of elementary functions. This process stops when removing a certain elementary function consistently increases the value of wAIC. As shown in Fig.\ 1e, each curve in a plot at the left column represents the information inconsistency vs. model complexity for one topological sample. And we find that, indeed, the joint operation with wAIC and topological sampling are helpful for inference from noisy and incomplete data (Fig.\ 5(k,l)).

The final sets of elementary functions and their coefficients $\hat{\boldsymbol{\xi}}_F$ and $\hat{\boldsymbol{\xi}}_G$ compose the forms $\hat{\textbf{F}}$ and $\hat{\textbf{G}}$, leading to the successful inference of network dynamics described by Eq.\ (\ref{EquDyn}). Indeed, as demonstrated in Fig.\ 1f, the trajectory generated by the inferred dynamical equation agrees well with the numerical simulations of the true network dynamics. It is worth noting that the ground truth, \textit{i.e.}\ the form of the true equation, keeps unknown during the whole procedure and is only used to assess the accuracy of the final inferred results, hence our approach works in an autonomous unsupervised way.

\subsection*{wAIC}
The original Akaike's Information Criterion (AIC)\cite{akaike1974new} is a frequently used method to balance the fitting and the complexity of a model with respect to the observed data, defined as $\text{AIC}=n\log{\text{MSE}}+2p$, where $n$ is the number of observations, $\text{MSE}$ is mean squared error of regression result of the model, and $p$ is the number of variables. By AIC, one aims to select an optimal model that best fits the  observations with the fewest variables from the model candidates. However, we find that the original AIC does not work well in the inference problem we aim to solve in the present work. Hence we introduce a weighted version of AIC (\textit{i.e.} wAIC),
\begin{equation}
\text{wAIC}=\left\{
\begin{aligned}
w\cdot(n\log{\text{MSE}}+2p), (n\log{\text{MSE}}+2p)\geq 0, \\
(n\log{\text{MSE}}+2p)/w, (n\log{\text{MSE}}+2p) < 0,
\end{aligned}
\right.
\end{equation}
to balance the fitting accuracy and the model complexity, where $w$ is the inferred coefficient of a term from Phase \uppercase\expandafter{\romannumeral1}. A term with a larger $w$ inferred by Phase \uppercase\expandafter{\romannumeral1} is more likely to be able to capture the properties of the underlying unknown dynamics. Thus, multiplying $w$ or $1/w$ with AIC amplifies the impact of removing this term from the equation. The smaller the wAIC, the more consistent is the composition of the elementary functions with observed data and less important is this removed term.

To be specific, to evaluate the relevance of a term $i$, we remove this term from the equation inferred by Phase \uppercase\expandafter{\romannumeral1} and calculate the value of $\text{wAIC}_i$ of the new shorter equation (Supplementary Figure 2). We repeat this process to obtain the wAIC for each term. Then we sort these terms based on their wAIC values, and remove terms one by one with wAIC values from small to large. This operation gives a shorter equation at each step, and we calculate these shortened equations' AIC values. The optimal equation is determined at the tuning point where the curve starts consistently increasing, as indicated by pink stars in Fig.\ 1e.

\subsection*{Topological sampling}
We perform topological sampling in phase \uppercase\expandafter{\romannumeral2} as follows. We randomly choose $S$ nodes from all $n$ nodes, and obtain the activities of these $S$ nodes' partial neighbors. Introducing the sampled ego structures and nodes' activities into libraries $L_F$ and $L_G$ allows us to construct the self and interaction matrices $\widetilde{\Theta}_F$ and $\widetilde{\Theta}_G$, and to further distill the elementary functions as well as their coefficients. We repeat the process to obtain $K$ sets of samples, average the coefficients of the elementary functions inferred from the $K$ sample sets. In the present work we set $S=10$ and $K=20$.

\subsection*{sMAPE}
The inference inaccuracy is quantified by symmetric mean absolute percentage error (sMAPE)\cite{flores1986pragmatic},
\begin{equation}
\text{sMAPE} = \frac{1}{m}\sum_{i=1}^{m}\frac{|I_i-R_i|}{(|I_i|+|R_i|)},
\end{equation}
where $m$ is the cardinal number of the set that contains both inferred and true elementary functions, $I_i$ and $R_i$ are inferred and true coefficients respectively. The range of sMAPE is $[0,1]$. The more accurate the inferred equation, the lower the value of sMAPE. Note that if an inferred elementary function should not exist in the true equation or a true elementary function is not successfully inferred, the value of sMAPE will significantly increase. Therefore, the metric sMAPE captures not only the errors of inferred coefficients but also the incorrectness of the inferred equation form.

\subsection*{Normalized Euclidean distance (NED)}
To evaluate the discrepancy between the inferred and true dynamics, we used the metric of normalized Euclidean distance (NED) that represents the distance between the two trajectories generated by the inferred and the true dynamical equations respectively. That is,
\begin{equation}
\text{NED}({x}_i,\hat{x}_i)=\frac{1}{D_{\text{max}}(T-t_0)}\sum_{t=t_0}^{T}\sqrt{(x_i(t)-\hat{x}_i(t))^2+(\dot{x}_i(t)-\dot{\hat{x}}_i(t))^2}.
\end{equation}
Here $x_i$ is the true trajectory and $\hat{x}_i$ is the trajectory generated by the inferred equation, $t_0$ and $T$ are the beginning and ending times respectively, and $D_{\text{max}}$ is the longest Euclidean distance between a pair of points of true trajectory.

\section*{Data availability}
The empirical networks data include \textit{C. elegans} connectome\cite{white1986structure,varshney2011structural,yan2017network}, the mushroom-body region of \textit{Drosophila}\cite{scheffer2020connectome}, the North-Europe power gird\cite{menck2014dead}, the U.S. power grid\cite{kunegis2013konect}, Advogato social network\cite{rossi2015network} retrieved from website \href{http://networkrepository.com}{http://networkrepository.com}, and the worldwide airline network retrieved from OpenFights data (\href{https://openflights.org/data.html}{https://openflights.org/data.html}). The empirical data of epidemic spreading include daily reported numbers of H1N1 and SARS cases available at Kaggle (\href{https://www.kaggle.com/lnunes/a-brief-comparative-study-of-epidemics/data}{https://www.kaggle.com/lnunes/a-brief-comparative-study-of-epidemics/data}), and the daily reported numbers of COVID-19 cases\cite{dong2020interactive}. Source Data are available at the Code Ocean capsule\cite{codelink}.

\section*{Code availability}
All source codes are publicly available at the Code Ocean capsule\cite{codelink}.

\section*{Acknowledgements}
TTG and GY are supported by the National Key Research and Development Program of China (grant no. 2021ZD0204500), National Natural Science Foundation of China (grant no. 12161141016 and 11875043), Shanghai Municipal Science and Technology Major Project (grant no. 2021SHZDZX0100), Shanghai Municipal Commission of Science and Technology Project (grant no. 18ZR1442000 and 19511132101), and Fundamental Research Funds for the Central Universities. The authors are also grateful for the helpful discussion with Prof. Baruch Barzel, Dr. Jack Moore, Dr. Xiaolei Ru and Mr. Tongyu Li.

\section*{Author contributions}
GY conceived the research, GY and TTG designed the research, TTG performed the research, TTG and GY analyzed the results, GY and TTG wrote this manuscript.

\section*{Competing interests}
The authors declare no competing interests.


\clearpage
\begin{figure*}[t]
	\centering
	\includegraphics[width=\textwidth]{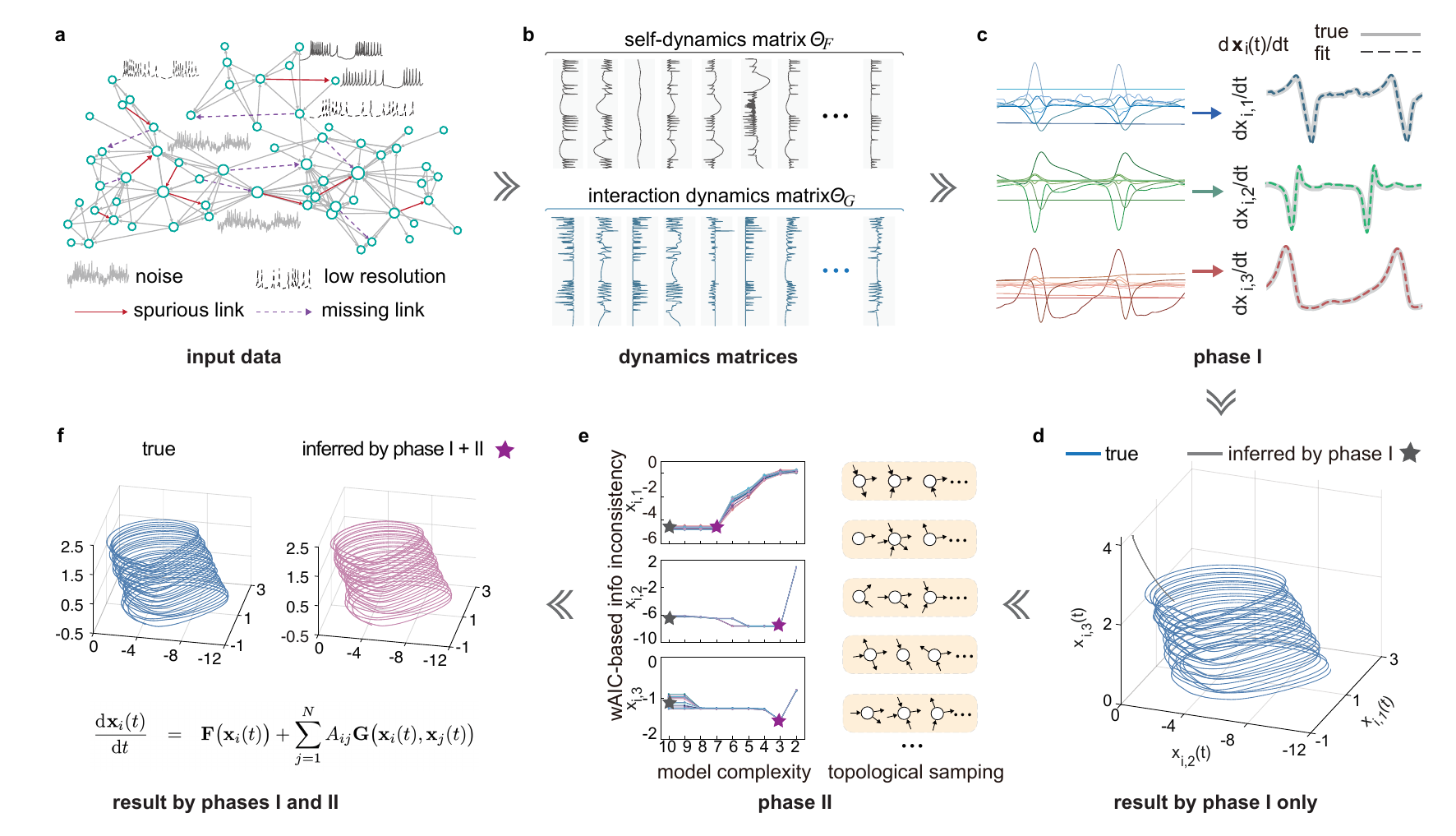}
	\caption{\textbf{Overview of the two-phase inference approach.} \textbf{a}. Observation data of network topology $A_{ij}$, including spurious and missing links, and low-resolution and noisy data of nodal activities $\textbf{x}_i(t)$. \textbf{b}. Mapping the normalized observation data into two matrices $\Theta_F$ and $\Theta_G$ that represent the time-varying patterns of elementary functions. \textbf{c}. Phase \uppercase\expandafter{\romannumeral1} that narrows down the model space by identifying several leading elementary functions through global regression for each dimension of $\dot{\textbf{x}}_i(t)$. \textbf{d}. Comparison of trajectories generated respectively by the true network dynamics and by the dynamical equation inferred by Phase \uppercase\expandafter{\romannumeral1} alone. \textbf{e}. Phase \uppercase\expandafter{\romannumeral2} that performs local fine-tuning, by using topological sampling and weighted Akaike's information criterion (wAIC), to further determine the optimal number (indicated by pink stars) of elementary functions for $\hat{\textbf{F}}(\textbf{x}_i(t))$ and $\hat{\textbf{G}}(\textbf{x}_i(t),\textbf{x}_j(t))$. \textbf{f}. Comparison of trajectories generated respectively by the true and the inferred dynamical equations. The example illustrated in (\textbf{c}-\textbf{f}) is Hindmarsh-Rose neuronal dynamics on a directed BA network with size $n=100$, average degree $\langle k\rangle=5$.}.
	\label{fig1}
\end{figure*}

\clearpage
\begin{figure}
	\centering
	\includegraphics[width=\textwidth]{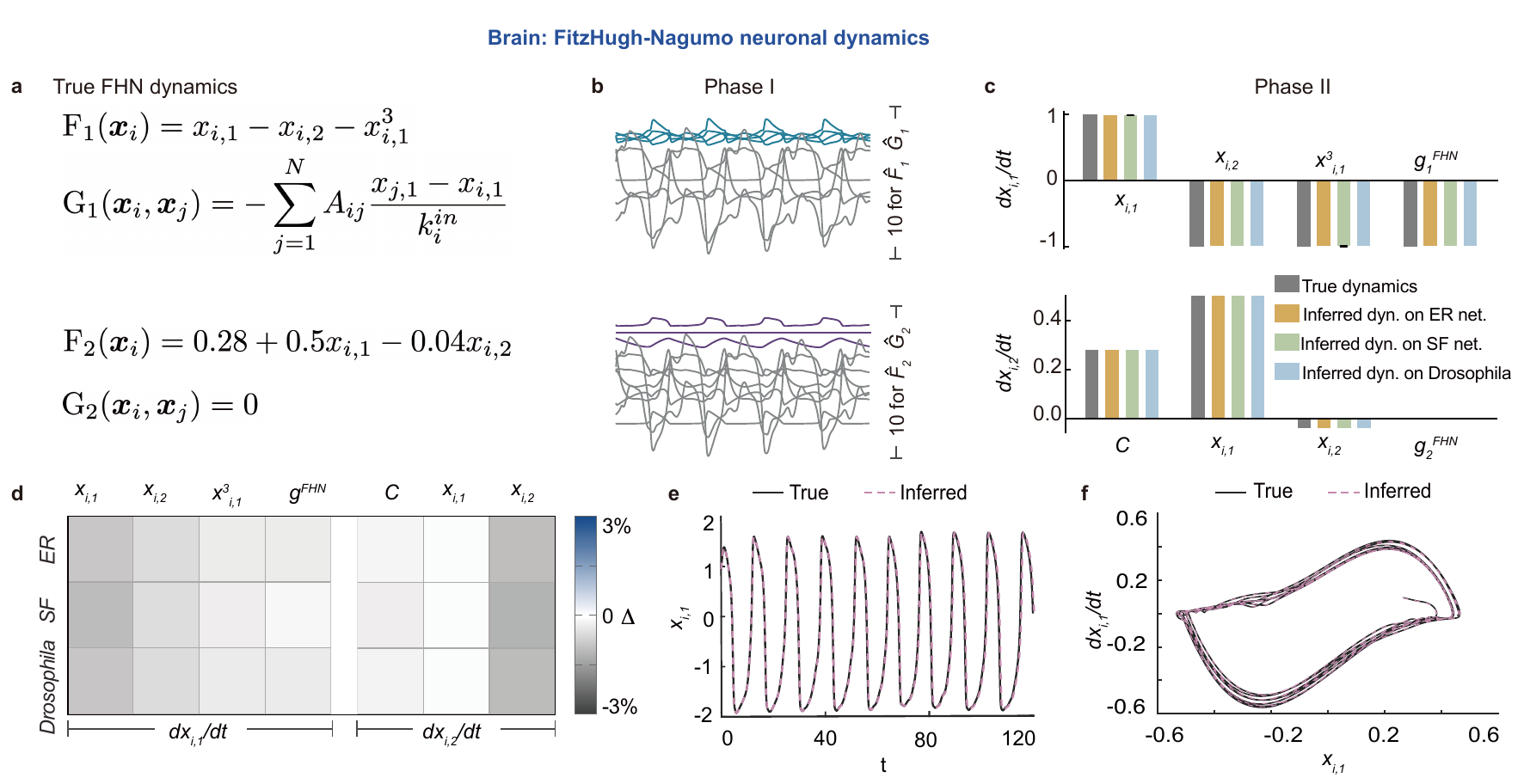}
	\centering
\caption{
\textbf{Inferring FitzHugh-Nagumo neuronal network dynamics on synthetic and real topologies.} \textbf{a}. True FitzHugh-Nagumo (FHN) dynamics used to simulate nodes' activities data on various topologies. $\text{F}_d$ and $\text{G}_d$ are self- and interaction dynamics of the $d$-th dimension respectively, $x_{i,d}$ is $d$-th dimension's state of node $i$, and $x_{i,d}^p$ is the polynomial with order $p$. \textbf{b}. Ten leading elementary functions identified by Phase \uppercase\expandafter{\romannumeral1} for each dimension. \textbf{c}. The necessary elementary functions and their coefficients further inferred through Phase \uppercase\expandafter{\romannumeral2} on two synthetic (directed ER and undirected SF) and one empirical (\textit{Drosophila} mushroom body) networks, where $g^{\text{FHN}}$ denotes the term $(x_j-x_i)/k_i^{\text{in}}$. \textbf{d}. Relative errors $\Delta$ of the inferred elementary functions and their coefficients. Note that the elementary functions ruled out from $\Theta_F$ and $\Theta_G$ by our approach (\textit{i.e.}\ whose coefficients are inferred as zero) are not shown. \textbf{e,f}. Nodes' activites and trajectories generated respectively by the true and the inferred equations.}
\label{fig2}
\end{figure}

\clearpage
\begin{figure}
	\centering
	\includegraphics[width=\textwidth]{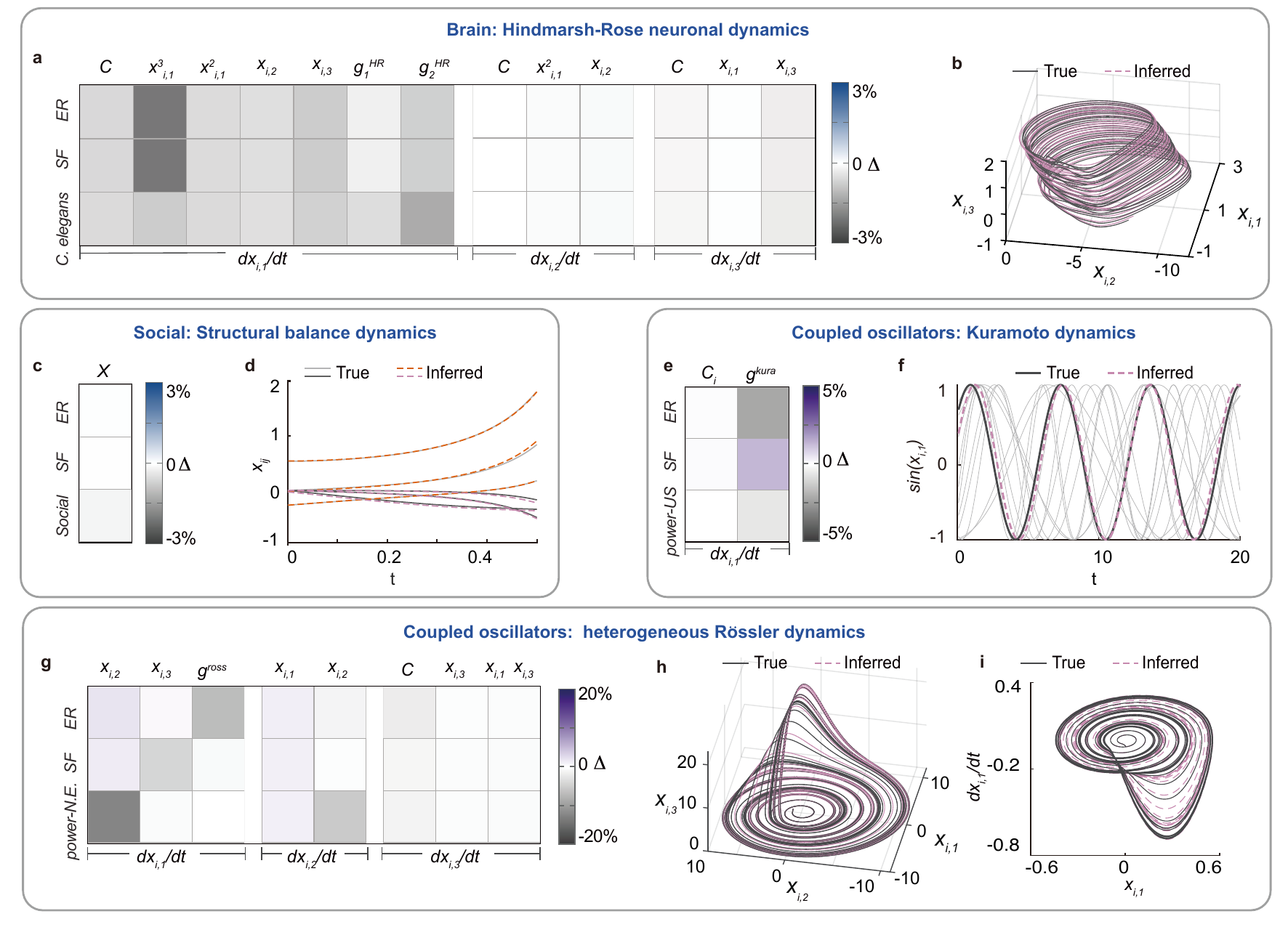}
	\centering
\caption{\textbf{Inference accuracy for other four typical nonlinear network dynamics.} \textbf{a,b}. Similar to Fig.\ 2 but for inferring HR neuronal dynamics, where the interaction dynamics $G(\boldsymbol{x}_i,\boldsymbol{x}_j)$ are composed of $g_1^{\text{HR}}\equiv 1/(1+e^{10(x_j-1)})$ and $g_2^{\text{HR}}\equiv x_i/(1+e^{10(x_j-1)})$. \textbf{c,d}. Relative errors and six edges' activities of the inferred edge dynamics of social balance. \textbf{e-i}. Relative errors of the inferred effective equations for network dynamics of Kuramoto model and coupled R\"{o}ssler oscillators. In both cases the self-dynamics are heterogeneous, \textit{i.e.}\ the intrinsic frequency of each node is not identical but follows a normal distribution $\mathcal{N}(1,\sigma)$ with $\sigma=0.1$. Grey curves represent the activity of individual nodes and black curves represents the averaging activity of systems. Symbols $g^{\text{kura}}$ and $g^{\text{ross}}$ denote terms $\sin{(x_j-x_i)}$ and $(x_j-x_i)$ respectively. The details of these dynamics and empirical networks are shown in Supplementary Tables 3 and 4.}
	\label{fig3}
\end{figure}

\clearpage
\begin{figure*}[t]
	\centering
	\includegraphics[width=\textwidth]{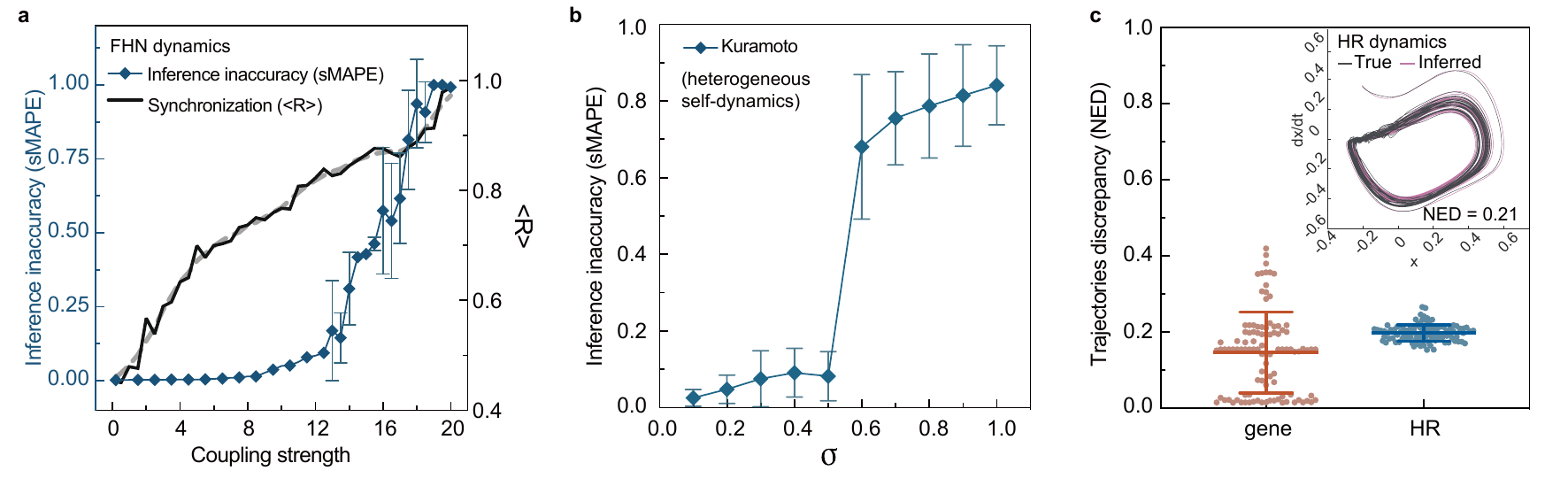}
	\caption{\textbf{Inferrability of network dynamics.} \textbf{a}. Inference inaccuracy represented by sMAPE and synchronization represented by order parameter $\langle R\rangle$ vs.\ coupling strength between nodes. \textbf{b}. Inaccuracy of inferred effective equation for Kuramoto network dynamics where the natural frequency $\omega$ of each node follows a normal distribution $\mathcal{N}(1,\sigma)$. Larger $\sigma$ indicates higher dynamical heterogeneity. \textbf{c}. Normalized Euclidean distance (NED, see Methods) when some true elementary functions were deliberately removed from libraries $L_F$ and $L_G$. The error bars represent the mean $\pm$ the standard deviation, and the sample size is 100. The networks are scale-free with size $n=100$ and average degree $\langle k\rangle=5.0$. Simulation details are shown in Supplementary Table 4.}
	\label{fig4}
\end{figure*}

\clearpage
\begin{figure*}[h!]
	\centering
	\includegraphics[width=\textwidth]{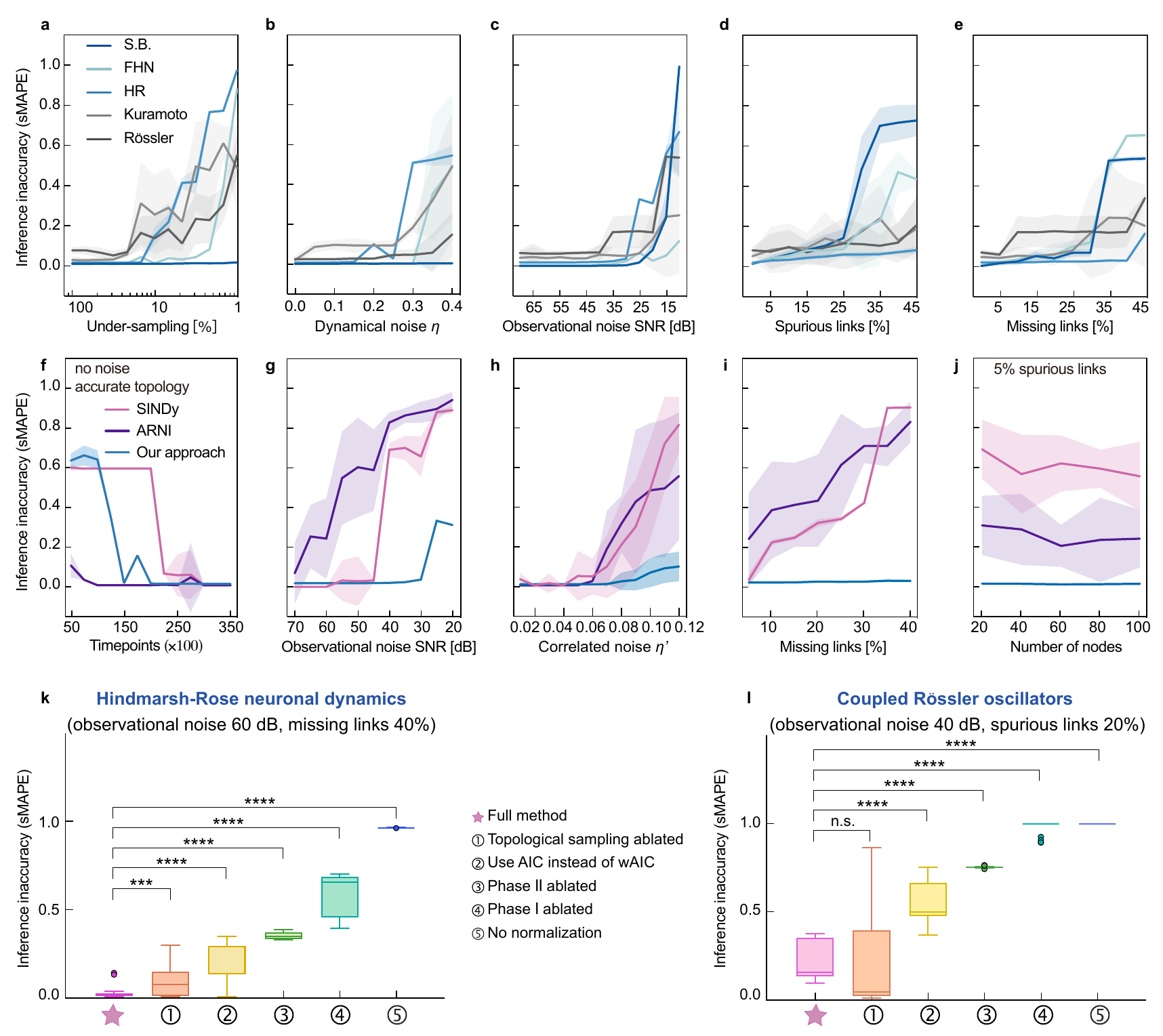}
	\caption{\textbf{Inference robustness against incompleteness and noises.} \textbf{a-e}. Inference inaccuracies sMAPE when the nodes' activities data are low-resolution, have dynamical noises (Gaussian white noise with intensity $\eta$) or observational noises (intensity quantified by signal-to-noise ratio SNR) or when the topologies data have spurious and missing links. \textbf{f-j}. Comparisons of inference inaccuracies between SINDy, ARNI and our approach for inferring HR neuronal network dynamics, with varying amount of time-points, observational noise, correlated dynamical noise, missing links, and different network sizes. Simulation details are shown in Supplementary Table 4. \textbf{k, l}. Comparison results of ablation studies. The box-whisker plots are visualized with the Tukey method (\textit{i.e.} the box represents the interquartile range (IQR) and the line in the box shows the median, with whiskers that extend $1.5$ times the IQR from the box edges; The outliers are also plotted) and the sample size is $20$. Five ablation studies were performed: \textcircled{1} removing topological sampling, \textcircled{2} using original AIC instead of wAIC, \textcircled{3} removing Phase \uppercase\expandafter{\romannumeral2}, \textcircled{4} removing Phase \uppercase\expandafter{\romannumeral1}, or \textcircled{5} without normalization to $\Theta_F$ and $\Theta_G$. Statistical significance is obtained through multiple Mann-Whitney tests. Three or four asterisks indicate $p$-value $<10^{-3}$ or $10^{-4}$, and n.s.\ means not significant.}
	\label{fig5}
\end{figure*}

\clearpage
\begin{figure*}[h!]
	\centering
	\includegraphics[width=\textwidth]{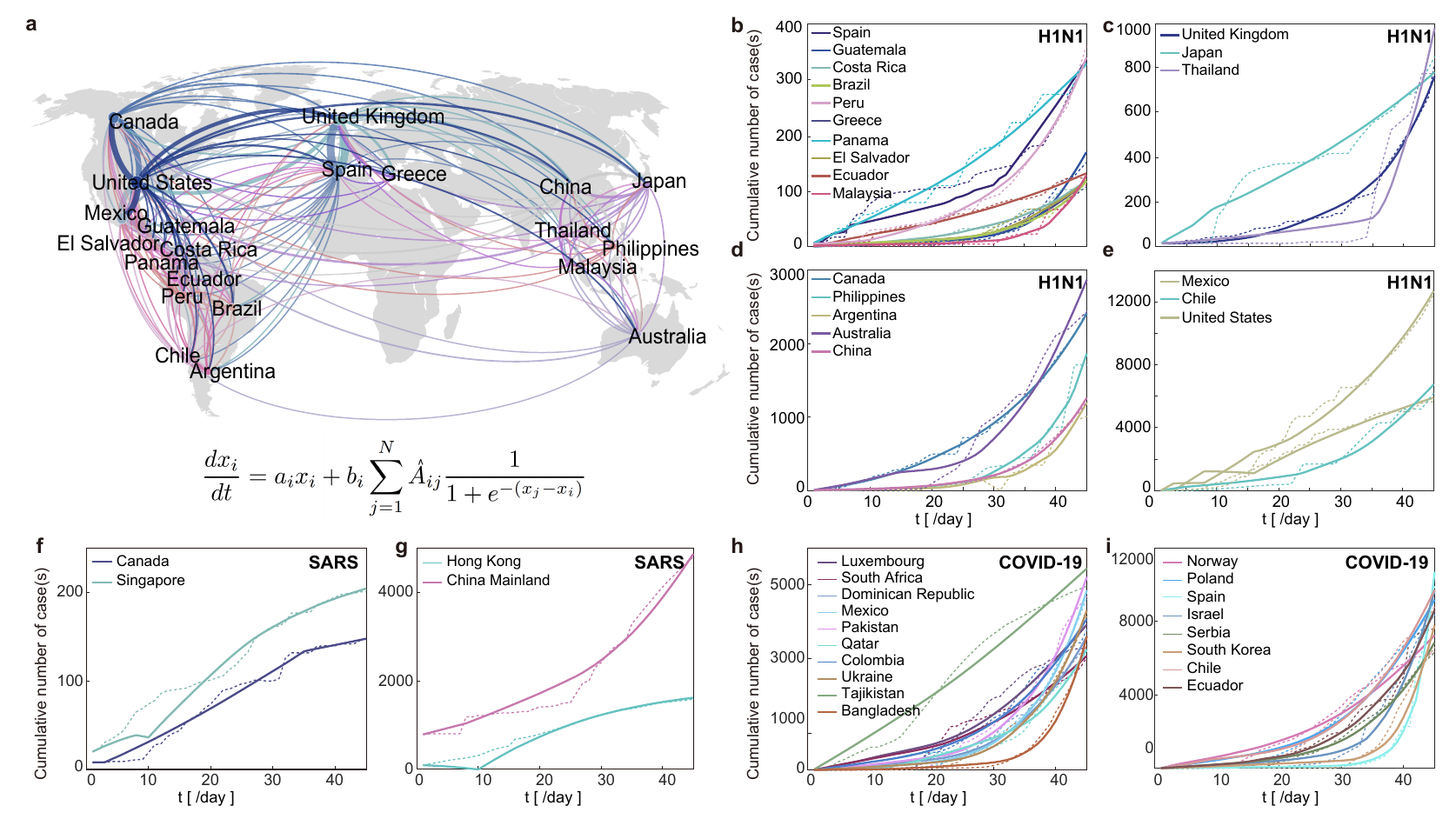}
	\caption{\textbf{Inference of early spreading dynamics from empirical data.} \textbf{a}. Worldwide airline network (partial) used for the inference. Each node represents a country or region, and line thickness represents the amount of passenger flow. The form describes the dynamical equation inferred by the two-phase approach. \textbf{b-e}. Comparisons between the empirical cumulative number of H1N1 cases for different nodes (dashed lines) and the cumulative number generated by the inferred equation (solid lines). For better visualization, the comparisons are displayed in four plots (from \textbf{b} to \textbf{e}) and the dates when the first case is reported for each node are all shifted to the first day ($t=1$). \textbf{f-i}. Comparisons between the empirical and the inferred cumulative numbers in different nodes for SARS (\textbf{f, g}) and COVID-19 (\textbf{h, i}).}
	\label{fig6}
\end{figure*}

\newpage
\renewcommand{\theequation}{S\arabic{equation}}
\renewcommand\tablename{Supplementary Table}
\renewcommand\figurename{Supplementary Figure}
\renewcommand{\citenumfont}[1]{[S#1]}
\renewcommand{\bibnumfmt}[1]{[S#1]}
\setcounter{figure}{0}

\begin{center}
{\Large \textit{Supplementary Information}} \\ \vspace{2mm} {\large \textbf{Autonomous inference of complex network dynamics from incomplete and noisy data}}
\vspace{30mm}

\tableofcontents
\end{center}

\newpage
\section{Approach}
\subsection{Function libraries}

We construct two comprehensive libraries, $L_F$ and $L_G$, for self- and interaction dynamics respectively, including polynomial, trigonometric, exponential, fractional, rescaling, and various activation functions as listed in Supplementary Tables \ref{self-matrix} and \ref{interaction-matrix}.

\begin{table}[!ht]
  \begin{center}
    \caption{Library $L_F$ for self-dynamics}
    \label{self-matrix}
    \begin{tabular}{c|c}
      \toprule
      \hline
      \textbf{Functions} &  $\boldsymbol{\Theta}_F=L_F(\textbf{x}_i)$\\
      \hline
      \textbf{Polynomial} & $\textbf{x}_i$, $\textbf{x}_i^2$, $\textbf{x}_i^3$, $\cdots$ \\
      \hline
      \textbf{Trigonometric} & $\sin({\textbf{x}_i})$, $\cos({\textbf{x}_i})$, $\tan({\textbf{x}_i})$  \\
      \hline
      \textbf{Exponential} & $e^{\textbf{x}_i}$ \\
      \hline
      \textbf{Fractional} & $\frac{1}{\textbf{x}_i}$\\
      \hline
      \textbf{Activation} & $\{\text{sigmoid}(\textbf{x}_i)\}^*$, $\tanh{(\textbf{x}_i)}$, $\{\frac{\textbf{x}_i^\gamma}{\textbf{x}_i^\gamma+1}\}^\dagger$\\
      \hline
      \textbf{Rescaling} & $\frac{\textbf{x}_i}{k_i^{\text{in}}}$\\
      \hline
      \bottomrule
    \end{tabular}
  \end{center}
\end{table}

\begin{table}[!ht]
  \begin{center}
    \caption{Library $L_G$ for interaction dynamics}
    \label{interaction-matrix}
    \begin{tabular}{c|c|c|c|c}
      \toprule
      \hline
      \textbf{Functions} & $\boldsymbol{\Theta}_G=L_G(\textbf{x}_j)$ & $\boldsymbol{\Theta}_G=L_G(\textbf{x}_i\textbf{x}_j)$ &$\boldsymbol{\Theta}_G=L_G(\textbf{x}_j-\textbf{x}_i)$&  $\boldsymbol{\Theta}_G=\textbf{x}_iL_G(\textbf{x}_j)$ \\
      \hline
      \textbf{Polynomial} & $\textbf{x}_j$, $\textbf{x}_j^2$, $\cdots$ & $\textbf{x}_i\textbf{x}_j$, $(\textbf{x}_i\textbf{x}_j)^2$, $\cdots$ & $\textbf{x}_j-\textbf{x}_i$, $(\textbf{x}_j-\textbf{x}_i)^2$, $\cdots$ &\\
      \hline
      \textbf{Trigonometric} & \makecell[c]{$\sin({\textbf{x}_j})$,\\ $\cos({\textbf{x}_j})$, $\cdots$} & \makecell[c]{$\sin({\textbf{x}_i\textbf{x}_j})$,\\ $\cos({\textbf{x}_i\textbf{x}_j})$, $\cdots$} & \makecell[c]{$\sin({\textbf{x}_j-\textbf{x}_i})$,\\ $\cos({\textbf{x}_j-\textbf{x}_i})$, $\cdots$}& \makecell[c]{$\textbf{x}_i\sin({\textbf{x}_j})$,\\ $\textbf{x}_i \cos(\textbf{x}_j)$, $\cdots$} \\
      \hline
      \textbf{Exponential} & $e^{\textbf{x}_j}$ & $e^{\textbf{x}_i\textbf{x}_j}$ & $e^{\textbf{x}_j-\textbf{x}_i}$&$\textbf{x}_i e^{\textbf{x}_j}$\\
      \hline
      \textbf{Fractional} & $\frac{1}{\textbf{x}_j}$ & $\frac{1}{\textbf{x}_i\textbf{x}_j}$ & $\frac{1}{\textbf{x}_j-\textbf{x}_i}$& $\frac{\textbf{x}_i}{\textbf{x}_j}$ \\
      \hline
      \textbf{Activation} & \makecell[c]{$\{\text{sigmoid}(\textbf{x}_j)\}^*$,\\ $\tanh{(\textbf{x}_j)}$,\\$\{\frac{\textbf{x}_j^\gamma}{\textbf{x}_j^\gamma+1}\}^\dagger$} & \makecell[c]{$\{\text{sigmoid}(\textbf{x}_i\textbf{x}_j)\}^*$,\\ $\tanh{(\textbf{x}_i\textbf{x}_j)}$,\\$\{\frac{(\textbf{x}_i\textbf{x}_j)^\gamma}{(\textbf{x}_i\textbf{x}_j)^\gamma+1}\}^\dagger$} & \makecell[c]{$\{\text{sigmoid}(\textbf{x}_j-\textbf{x}_i)\}^*$,\\ $\tanh{(\textbf{x}_j-\textbf{x}_i)}$,\\$\{\frac{(\textbf{x}_j-\textbf{x}_i)^\gamma}{(\textbf{x}_j-\textbf{x}_i)^\gamma+1}\}^\dagger$}&\makecell[c]{$\{\textbf{x}_i\text{sigmoid}(\textbf{x}_j)\}^*$,\\ $\textbf{x}_i\tanh{ (\textbf{x}_j)}$,\\$\{\frac{\textbf{x}_i\textbf{x}_j^\gamma}{\textbf{x}_j^\gamma+1}\}^\dagger$} \\
      \hline
      \textbf{Rescaling} & $\frac{\textbf{x}_j}{k_i^{\text{in}}}$ & $\frac{\textbf{x}_i\textbf{x}_j}{k_i^{\text{in}}}$ & $\frac{\textbf{x}_j-\textbf{x}_i}{k_i^{\text{in}}}$&$\frac{\textbf{x}_i\textbf{x}_j}{k_i^{\text{in}}}$\\
      \hline
      \bottomrule
    \end{tabular}
  \end{center}
\end{table}
\footnotetext[1]{sigmoid: indicates a variety of sigmoid functions $\frac{1}{1+e^{-\alpha(x-\beta)}}$, where $\alpha \in [1:1:10]$ and $\beta \in [0:1:10]$.}
 \footnotetext[2]{$\gamma \in [0:1:10]$.}

\begin{figure}[!ht]
\centering
\includegraphics[width=0.95\linewidth]{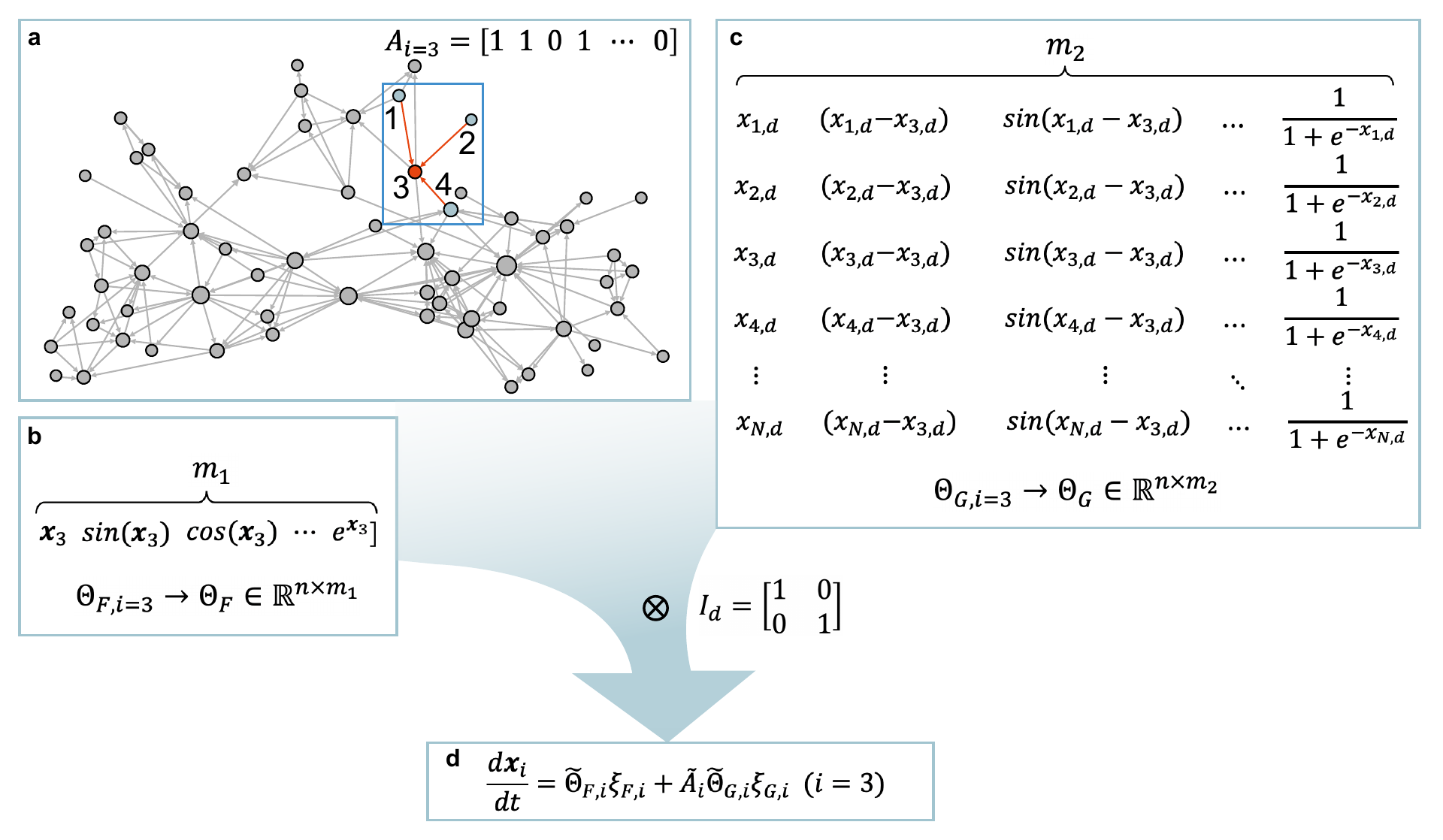}
\label{Schematic}
\caption{An example for the construction of matrices $\Theta_F$ and $\Theta_G$. \textbf{a}. A network, and the adjacency vector $A_{i=3}$ of node $3$. \textbf{b}. Constructing self-dynamics matrix $\Theta_{F,i=3}$ by mapping $\textbf{x}_3(t)$ to library $L_F$ that contains $m_1$ elementary functions. \textbf{c}. Constructing interaction dynamics matrix $\Theta_{G,i=3}$ by mapping $\textbf{x}_3(t)$ and $\{\textbf{x}_j(t)\}$, where $j$ is in the set of node $i$'s neighbors, to library $L_G$ that contains $m_2$ interaction elementary functions. Through the Kronecker product $I_d$ of the identity matrix and $\Theta_{F,i=3}$ (or $\Theta_{G,i=3}$), we obtain matrices $\widetilde{\Theta}_{F,i=3}$ and $\widetilde{\Theta}_{G,i=3}$ capturing the whole $d$-dimensional nodes' activities on all elementary functions. Hence the optimization is to search for $\xi_{F,i=3}$ and $\xi_{G,i=3}$ that best solve the equation exhibited in \textbf{d}.}
\centering
\end{figure}

\begin{figure}[!ht]
    \centering
    \includegraphics[width=0.45\linewidth]{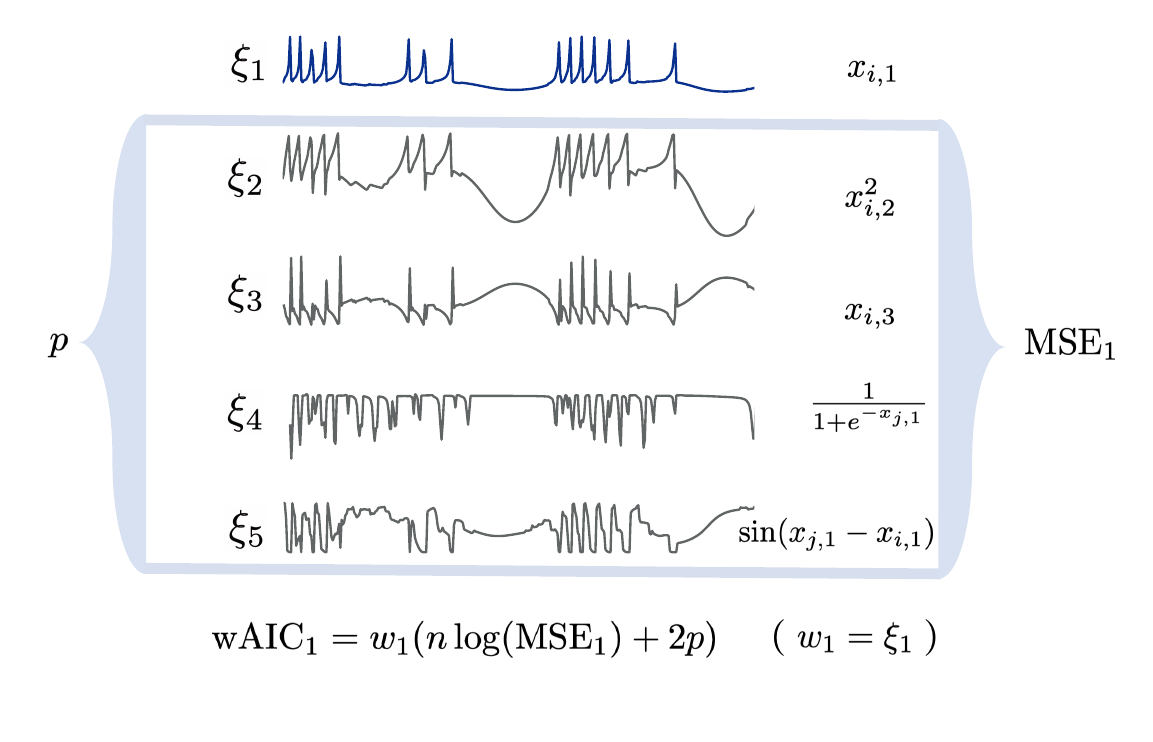}
    \label{wAIC}
    \caption{An example for the calculation of wAIC. To evaluate the relevance of the term $x_{i,1}$ to the true dynamics, we remove this term from the set of leading terms inferred by Phase \uppercase\expandafter{\romannumeral1} and then calculate the $\text{MSE}_1$ between the dynamics composed of the remaining $p$ terms and the observation data. Then we obtain the value of $\text{wAIC}_1$ that represents the relevance of this removed term (the higher the more relevant).}
    \centering
\end{figure}

\subsection{Pseudocode}

\textbf{Phase \uppercase\expandafter{\romannumeral1}}. As shown in Supplementary Figure 3, the values of the terms (\textit{i.e.} elementary functions) in libraries $L_F$ and $L_G$ can span several orders of magnitude. Such a wide spread has detrimental effects on evaluating the relevance of each term to the true unknown dynamics. For example, if a term is inherently low-valued, a regression method will probably assign a relatively large coefficient to it, which might indicate the term is more relevant. However, this is not always true since it could be caused by the fact that the scale of this term is inherently smaller than that of other terms. To remove such detrimental effect, we hence do not use the original data but rather perform normalization for the derivatives $\Dot{\textbf{x}}$ and each column of the matrices $\Theta_F$ and $\Theta_G$. To be specific, the normalization of the $i$-th column of the matrix $\Theta$ is
\begin{equation}
    \Vert \Theta_i\Vert = \frac{\Theta_i}{\sum_{k=1}^{n\times T}\Theta_{i,k}},
\end{equation}
and the normalization of the $d$-th dimension of the derivative is
\begin{equation*}
    \Vert \Dot{\textbf{x}}_d\Vert=\frac{\Dot{\textbf{x}}_d}{\sum_{k=1}^{n\times T}\Dot{\textbf{x}}_{d,k}}.
\end{equation*}

Therefore Phase \uppercase\expandafter{\romannumeral1} aims to obtain the coefficients $\xi_m$ in
\begin{equation}
    \Dot{\textbf{x}}_d=\sum_{m=1}^{m_1+m_2}\Theta_m (\xi_m\cdot\frac{\sum_{k=1}^{n\times T}\Dot{\textbf{x}}_{d,k}}{\sum_{k=1}^{n\times T}\Theta_{m,k}}),
\end{equation}
where $m_1$ and $m_2$ are the numbers of terms in libraries $L_F$ and $L_G$ respectively. The pseudocode of Phase \uppercase\expandafter{\romannumeral1} is described in Algorithm 1.
\begin{figure}[!ht]
\centering
\includegraphics[width=0.6\linewidth]{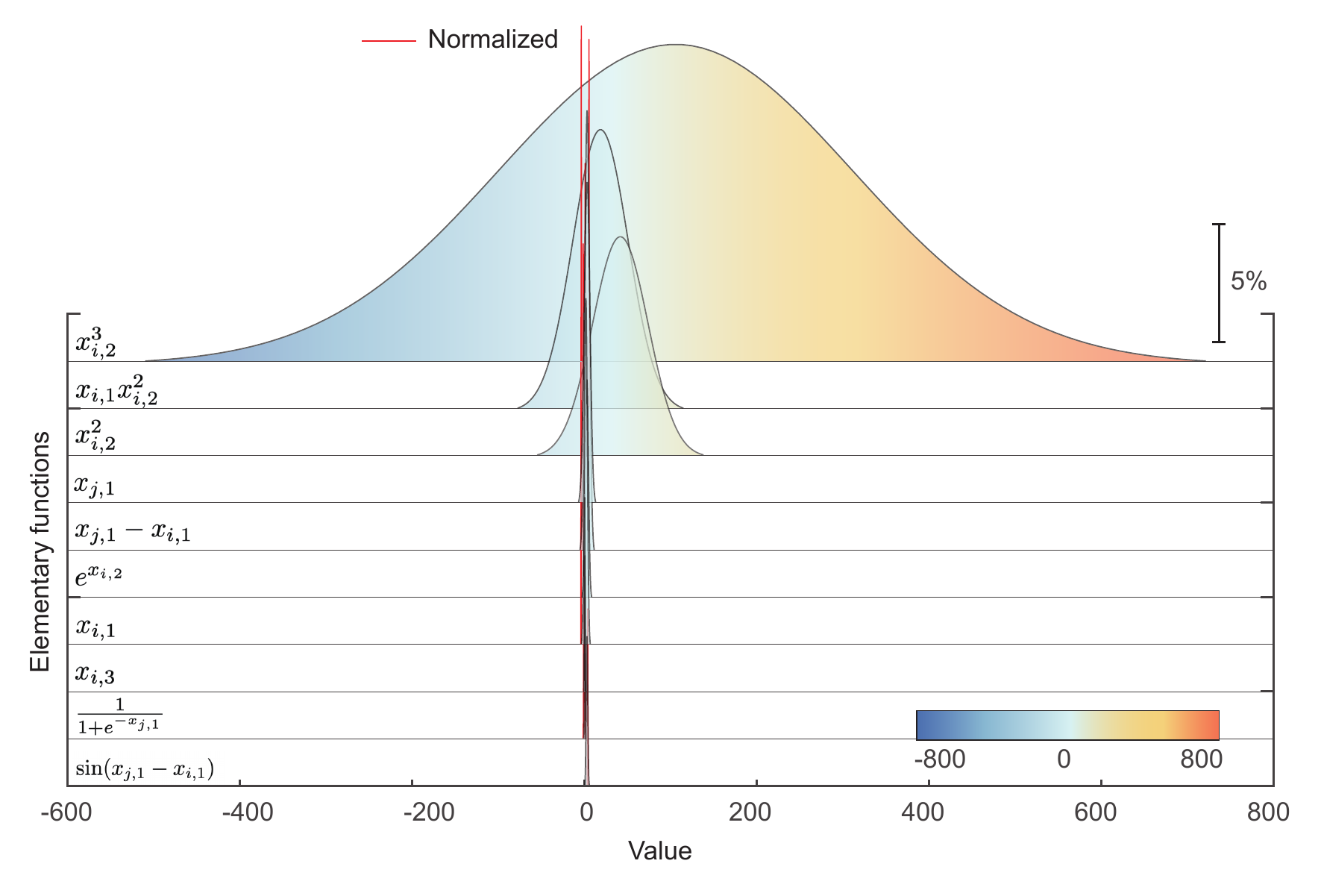}
\label{Normalization comparison}
\caption{The distribution of the values of typical elementary functions with or without normalization. The terms $x_{i,2}^3$, $x_{i,1}x_{i,2}^2$ and $x_{i,2}^2$ can span several orders of magnitude.}
\centering
\end{figure}

\begin{algorithm}[!ht]
  \caption{Phase \uppercase\expandafter{\romannumeral1}}
  \label{alg::phase1}
  \begin{algorithmic}[1]
    \Require
      $A$: network topology

      $\Dot{\textbf{x}}$: numerical derivatives

      $\widetilde{\Theta}_F,\widetilde{\Theta}_G$: libraries' matrices, for self- and interaction dynamics respectively

      $\kappa$: the number of leading terms to be identified

    \Ensure
      ${L}_F^*$, ${L}_G^*$: reduced libraries for self- and interaction dynamics respectively

      \hspace*{3mm}$\boldsymbol{w}$: weight vector

    \State \# \textit{data normalization}
    \State normalizing of each column of $\Dot{\textbf{x}}$, $\widetilde{\Theta}_F$ and $\widetilde{\Theta}_G$;

    \State \# \textit{global regression}
    \State minimizing of formula (3) in main paper via lasso to  obtain $\boldsymbol{\xi}_F^*$, $\boldsymbol{\xi}_G^*$, and intercept value $c$;
    \State \# \textit{Judging whether the constant term $c$ should exist}
    \State Temporally construct two reduced libraries: one contains only the $\kappa$ leading terms with the largest absolute coefficients $\boldsymbol{\xi}^*$, the other contains the $\kappa$ leading terms and the constant term;
    \State compute $\text{AIC}_{\text{w.c.}}$ and $\text{AIC}_{\text{wo.c.}}$ for libraries with and without the constant term respectively.
    \State \textbf{if} {$\text{AIC}_{\text{w.c.}}-\text{AIC}_{\text{wo.c.}}> \beta_1\text{AIC}_{\text{w.c.}}$}  \hspace*{1cm}\# \textit{$\beta_1$: a threshold for the judgement}
    \State  \hspace*{5mm}the constant term is necessary;
    \State \textbf{else}
    \State  \hspace*{5mm}the constant term is unnecessary;
    \State \textbf{end}
  \end{algorithmic}
\end{algorithm}

By Phase \uppercase\expandafter{\romannumeral1}, the comprehensive libraries containing a large number of elementary functions that potentially compose the self- and interaction dynamics are significantly narrowed down to ${L}_F^*$ and ${L}_G
^*$. Because, for most systems of interest, the underlying dynamical equation often consists of only a few terms, the reduced libraries ${L}_F^*$ and ${L}_G^*$ contain top $\kappa$ terms with the largest absolute coefficients described by the weight vector $\boldsymbol{w}$. To determine whether the hidden equation should contain a constant term, we compare the AIC values of the equations with and without constant. If the AIC value of the equation with constant is far smaller than that without constant, the hidden equation probably has a constant term. In the present paper we set $\kappa=10$, \textit{i.e.} obtain top ten leading elementary functions for ${L}_F^*$ and ${L}_G^*$ by algorithm \uppercase\expandafter{\romannumeral1}.

Note that the hyper-parameter $\lambda$ in Phase \uppercase\expandafter{\romannumeral1} is automatically selected by $5$-fold cross validation, a widely-used trick to estimate the most appropriate value of hyper-parameter. Specifically, to choose the best $\lambda$ from a set $\{\lambda_1,\lambda_2,\cdots,\lambda_r\}$, we parameterize the loss function $(\mathcal{L}(\lambda_1),\mathcal{L}(\lambda_2),\cdots,$ $\mathcal{L}(\lambda_r))$, and randomly divide the nodes' activities data into five sets, with one set for test and the other four sets for training. We apply each value of $\lambda$ and calculate the corresponding score of lasso regression. Finally the hyper-parameter with the highest regression score is selected as the optimal value of $\lambda$\cite{scikit-learn}.

\textbf{Phase \uppercase\expandafter{\romannumeral2}}. Through Phase \uppercase\expandafter{\romannumeral1} we have obtained a reduced model space that represents the set of most relevant elementary functions. With the reduced libraries ${L}_F^*$, ${L}_G^*$ at hand, next we propose Phase \uppercase\expandafter{\romannumeral2} to infer a concise form as well as appropriate coefficients for the hidden dynamical equation. The procedure of Phase \uppercase\expandafter{\romannumeral2} has two key sub-steps, \textit{i.e.} topological sampling, and fine-tuning with wAIC (see Methods). The pseudocode of Phase \uppercase\expandafter{\romannumeral2} is described in Algorithm 2.
\begin{algorithm}[!ht]
  \caption{Phase \uppercase\expandafter{\romannumeral2}}
  \label{alg::phase2}
  \begin{algorithmic}[1]
    \Require
      $\Dot{\textbf{x}}$: numerical derivatives

      ${L}_F^*$, ${L}_G^*$: the reduced libraries by Phase \uppercase\expandafter{\romannumeral1}

      $\boldsymbol{w}$: weight vector obtained by Phase \uppercase\expandafter{\romannumeral1}


    \Ensure
      $\hat{L}_F$, $\hat{L}_G$: the final inferred terms for self- and interaction dynamics respectively

      \hspace*{3mm} $\hat{\boldsymbol{\xi}}_F$, $\hat{\boldsymbol{\xi}}_G$: the final inferred coefficients for self- and interaction dynamics respectively

    \State extract the reduced matrices $\hat{\Theta}_F$ and $\hat{\Theta}_G$ from $\widetilde{\Theta}_F$ and  $\widetilde{\Theta}_G$ based on ${L}_F^*$ and ${L}_G^*$;
    \State \# \textit{topological sampling}
    \State \textbf{for} $i=1$; $i<K$; $i++$ \hspace*{1.8cm} \# \textit{$K$: number of topological samples}
    \State  \hspace*{5mm}randomly chose $S$ nodes \hspace*{1.05cm} \# \textit{$S$: batch size}
    \State  \hspace*{5mm}merge the numerical derivatives and libraries matrices of $S$ nodes: $\Dot{\textbf{x}}^S$, $\hat{\Theta}_F^S$ and $\hat{\Theta}_G^S$;
    \State  \hspace*{5mm}\# \textit{assessment with wAIC}
    \State  \hspace*{5mm}\textbf{for} {$j=1$; $j<=\kappa$; $j++$}
    \State  \hspace*{10mm}remove the term $j$
    \State  \hspace*{10mm}compute $\text{AIC}_j$ for the combination of terms without term $j$;
    \State \hspace*{10mm}$\text{wAIC}_j\gets w_j\cdot \text{AIC}_j$;
    \State \hspace*{5mm}\textbf{end}

    \State \hspace*{5mm}\textbf {for} {$l=1$; $l<\kappa$; $l++$}
    \State \hspace*{10mm}remove $l$ terms that have smallest wAIC value;
    \State \hspace*{10mm}compute $\text{AIC}_{\kappa-l}$ for the combination of all the remaining terms;
    \State \hspace*{15mm}\textbf{if} {$\text{AIC}_{\kappa-l-1}-\text{AIC}_{\kappa-l}>\beta_2$} \hspace*{1cm}\# \textit{$\beta_2$: a threshold for the assessment}
    \State \hspace*{20mm}obtain the inferred elementary functions $\hat{L}_{F,i}$, $\hat{L}_{G,i}$ and their coefficients of the $i$-th topological sample by deleting $l$ terms;
    \State \hspace*{15mm}\textbf{end}
    \State \hspace*{5mm}\textbf{end}
    \State \textbf{end}
    \State obtain $\hat{L}_F$ and $\hat{L}_G$ by merging the identified terms of all $K$ topological samples;
    \State obtain $\hat{\boldsymbol{\xi}}_F$ and $\hat{\boldsymbol{\xi}}_G$ by averaging the inferred coefficients across all $K$ topological samples;
  \end{algorithmic}
\end{algorithm}
\section{Networks}
In the present paper we used both synthetic and empirical networks to validate our two-phase inference approach. The synthetic networks are generated by directed Erd\H{o}s-R\'{e}nyi (ER) model\cite{erdHos1960evolution}for random topologies, as well as Barab\'{a}si-Albert (BA) \cite{barabasi2003scale} and the static model\cite{goh87universal} for scale-free (SF) topologies. The empirical network datasets are obtained from neuronal, social and technological domains.

\subsection{Synthetic networks}
Erd\H{o}s-R\'{e}nyi (ER) networks are generated by \textit{erdos\_renyi\_graph(n, p)} in \textit{networkX}\cite{SciPyProceedings_11}, where $n=100$, $p=0.05$. The resulting average degree $\langle k \rangle=5.2$. For each link between node $i$ and $j$, we assign the link direction, \textit{i.e.} from $i$ to $j$, from $j$ to $i$, or a reciprocal connection between them, with equal probability.

Directed SF networks are generated by Barab\'{a}si-Albert model, \textit{i.e.} using \textit{barabasi\_albert\_graph} \textit{(n, m)} in \textit{networkX}\cite{SciPyProceedings_11}, where $n=100$, $m=5$. We set each link bidirectional, and then randomly delete a proportion of these unidirectional links while ensuring that the network is weakly-connected. Finally we obtain a directed network with average total-degree $\langle k \rangle=5.1$.

Directed scale-free networks are generated by the static model\cite{goh87universal}. The procedure is as follows: we assign a weight $\omega_i=i^{-\eta}$ to the node with index $i$, where $i=1, 2, \cdots, n$ and $0< \eta \leq 1$, then randomly select a pair of nodes $i$ and $j$ ($i\neq j$) with probability proportional to $\omega_i$ and $\omega_j$ respectively. If there the link from node $i$ to $j$ already exists, we randomly select another pair of nodes with proportional to their weights; Otherwise, we connect node $i$ to $j$. Repeat this process until the number of links reaches $mn$. Hence the average total-degree $\langle k \rangle=2m$, and the power-law exponent $\gamma=\frac{1+\eta}{\eta}$ which can be tuned by $\eta$.

\subsection{Empirical networks}

The empirical networks used in this work include \textit{C. elegans} connectome with 279 neurons\cite{white1986structureSI,varshney2011structuralSI,yan2017networkSI}, the mushroom-body region of \textit{Drosophila} with 1832 neurons\cite{scheffer2020connectomeSI}, the North-Europe (N.E.) power grid with 236 nodes\cite{menck2014deadSI}, the U.S. power grid with 4941 nodes\cite{kunegis2013konectSI} and a community in the weighted social network of Advogato with 623 nodes\cite{nr}. The properties of these empirical networks are described in Supplementary Table \ (\ref{network info}).
\begin{table}[!ht]
  \begin{center}
    \caption{Synthetic and empirical networks used in the paper. For each network, we show its type, name and reference, number of nodes $n$, directed or undirected, average degree $\langle k \rangle$, maximum ($k_{\text{max}}$) and minimal ($k_{\text{min}}$) degree, and degree heterogeneity $\frac{\langle k^2\rangle}{\langle k\rangle^2}$.}
    \label{network info}
    \resizebox{.95\columnwidth}{!}{
    \begin{tabular}{c c|c c c c c c}
      \toprule
      \hline
      type&name&n& dir/undir & $\langle K\rangle$ &$k_{\text{max}}$ &$k_{\text{min}}$ & $\frac{\langle k^2\rangle}{\langle k\rangle^2}$\\
      \hline
      \textcolor[RGB]{33,102,172}{\textbf{synthetic}}&SF (BA)\cite{barabasi2003scale} & 100&dir&5.1&40&3&1.5 \\

      &SF,($\gamma=2.5$)\cite{goh87universal} & 100&undir &2.6 &34 &1 &4.4 \\

      &ER\cite{erdHos1960evolution} & 100&dir &5.2 &21&4 &1.1 \\

      \hline
      \textcolor[RGB]{27,120,55}{\textbf{empirical}}&\textit{C. elegans}\cite{white1986structureSI,varshney2011structuralSI,yan2017networkSI}* & 279&dir &10.7 &137&2&1.7\\

      &\textit{Drosophila}\cite{scheffer2020connectomeSI}** &  1832&dir&6.6 &56 &1 &5.4 \\

      &North-Europe power grid\cite{menck2014deadSI} &236&undir &2.7 &13&1 &1.2 \\

      &U.S. power grid\cite{kunegis2013konectSI} & 4941&undir &2.7 &19 &1 &1.5\\

      &Advogato\cite{nr}*** & 623&dir &6.0 &251 &1 &4.4 \\
      \hline
      \bottomrule
    \end{tabular}}
  \end{center}
\end{table}

* Neuronal connectome of the nematode worm \textit{Caenorhabditis elegans}, which is defined anatomically at a cellular network as 2990 synaptic connections between 279 neurons.

** The cellular-level connectome of \textit{Drosophila} mushroom-body region that contains neurons, which project their axons within tracts resembling pairs of mushrooms, accessed via https://neuprint-examples.janelia.org.

*** A subgraph of the directed weighted network of trust among developers in Advogato platform. The edges can have positive or negative weights representing the amount of trust or distrust. We divide the whole network into communities and use a large community in this work.

\section{Dynamics and inferred results}
In this paper we demonstrated the effectiveness of our two-phase approach in inferring a wide range of complex network dynamics pertaining to brain, social and coupled nonlinear oscillators systems.
\subsection{Brain dynamics}
\subsubsection*{Hindmarsh-Rose dynamics}
Neurons exhibit spiking activities which is believed to be an essential component in information processing in brains\cite{rabinovich2006dynamicalSI}. To simulate such activities we employed the $3$-dimensional Hindmarsh-Rose (HR) model\cite{borges2018inference}, a moderately simplified version of the HodgkinšCHuxley\cite{rabinovich2006dynamicalSI} model. The true equations governing the HR network dynamics are
\begin{equation}
    \label{trueHR}
    \left\{
    \begin{aligned}
    \frac{dx_{i,1}}{dt}&=x_{i,2}-ax_{i,1}^3+bx_{i,1}^2-x_{i,3}+I_{ext}+\epsilon(V_{\text{syn}}-x_{i,1})\sum_{j=1}^{N}A_{ij}\mu(x_{j,1}), \\
    \frac{dx_{i,2}}{dt}&=c-ux_{i,1}^2-x_{i,2}, \\
    \frac{dx_{i,3}}{dt}&=r[s(x_{i,1}-x_0)-x_{i,3}],
    \end{aligned}
    \right.
\end{equation}
where the coupling term
\begin{equation}
    \label{HRcoup}
    \mu(x_{j,1})=\frac{1}{1+e^{[-\lambda(x_{j,1}-\Omega_{\text{syn}})]}}.
\end{equation}
Here $x_{i,1}$ is the membrane potential of neuron $i$, $x_{i,2}$ is the transport rate of sodium and potassium ions across the membrane through the ion channels, and $x_{i,3}$ is the adaptation current. The $x_{i,1}$ is a simplified notation for $x_{i,1}(t)$. Parameters $a=1$, $b=3$, $c=1$, $u=5$, $s=4$, $r=0.005$, $x_0=-1.6$, coupling strength $\epsilon=0.15$, $V_{\text{syn}}=2$, $\lambda=10$, $\Omega_{syn}=1$, and $I_{ext}$ is external current which is set to $3.24$ for all neurons. The coupling strength $\epsilon$ is tunable according to size and topology of networks (see sec. \uppercase\expandafter{\romannumeral4}).

Applying our approach to the neuronal activities data generated by HR dynamics on a directed scale-free network, we obtain the inferred equations typically as
\begin{equation}
    \label{inferHR}
    \left\{
    \begin{aligned}
    \frac{d\hat{x}_{i,1}}{dt}&=0.994x_{i,2}-0.993x_{i,1}^3+2.977x_{i,1}^2-0.977x_{i,3}+3.218\\
    &+\sum_{j=1}^{N}A_{ij}(0.296g_1^{HR}(x_{j,1})-0.146g_2^{HR}(x_{i,1},x_{j,1})), \\
    \frac{d\hat{x}_{i,2}}{dt}&=1.000-4.994x_{i,1}^2-0.999x_{i,2}, \\
    \frac{d\hat{x}_{i,3}}{dt}&=0.032+0.020x_{i,1}-0.005x_{i,3},
    \end{aligned}
    \right.
\end{equation}
where $g_1^{HR}=\frac{1}{1+e^{[-10(x_{i,1}-1)]}}$ and $g_2^{HR}=\frac{x_{j,1}}{1+e^{[-10(x_{i,1}-1)]}}$.
Note that all true elementary functions have been successfully inferred, and inaccuracies of their coefficients are lower than $3\%$ (as also shown in Fig.\ 3a in the main paper). The neuronal activities and trajectories generated by the true Eq. (\ref{trueHR}, \ref{HRcoup}) and the inferred Eq. (\ref{inferHR}) are shown in Supplementary Fig.\ 4.
\begin{figure*}[!ht]
	\centering
	\includegraphics[width=0.9\textwidth]{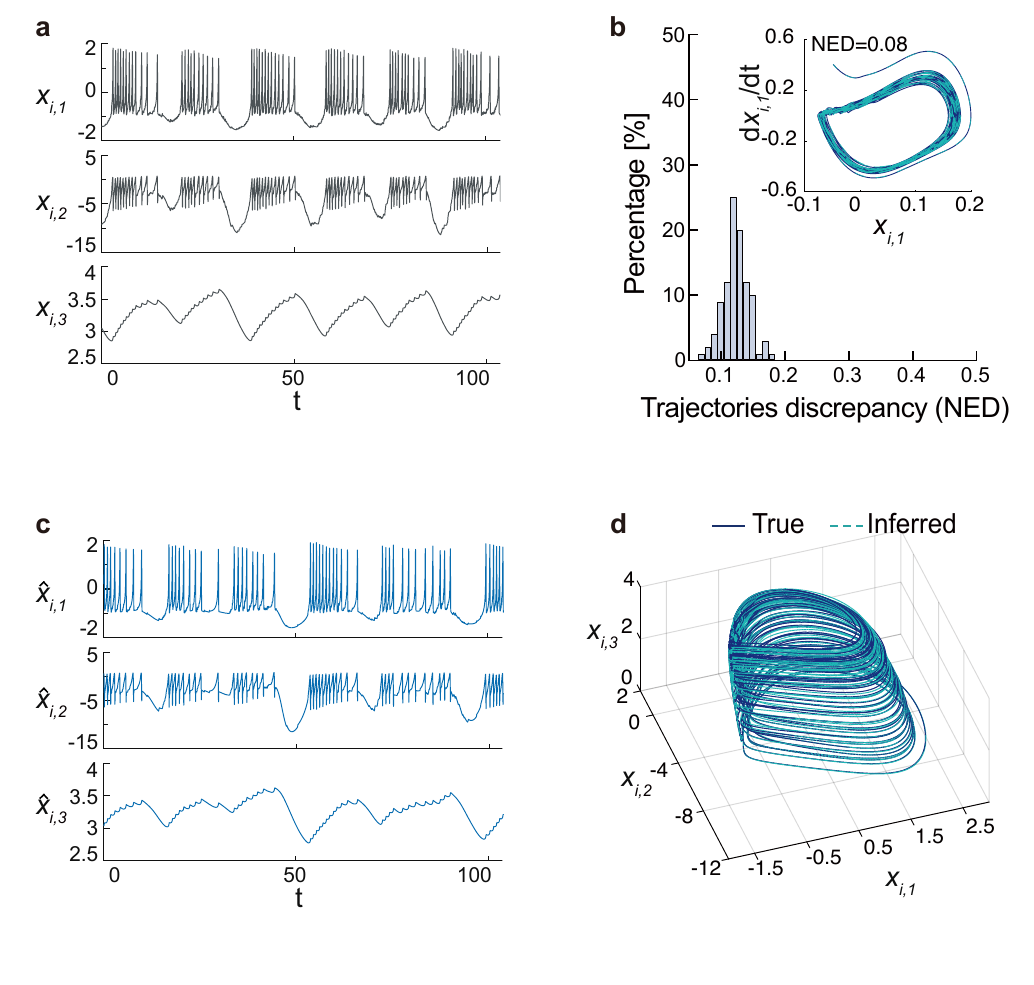}
	\caption{Neuronal activities and trajectories generated by the true HR network dynamics and the inferred dynamical equations.}
\end{figure*}

\subsubsection*{FitzHugh-Nagumo dynamics}

We also tested our approach by applying it to the neuronal activities data generated by FitzHugh-Nagumo (FHN) dynamics\cite{fitzhugh1961impulses} on various networks. The equations governing the FHN neuronal network dynamics are
\begin{equation}
    \label{trueFHN}
    \left\{
    \begin{aligned}
        \frac{dx_{i,1}}{dt}&=x_{i,1}-x_{i,1}^3-x_{i,2}-\epsilon\sum_{j=1}^{N}A_{ij}\frac{(x_{j,1}-x_{i,1})}{k_i^{in}},\\
        \frac{dx_{i,2}}{dt}&=a+bx_{i,1}+cx_{i,2}.
    \end{aligned}
    \right.
\end{equation}
The FHN dynamics capture the firing behaviors of neurons with two components. The first component represents the membrane potential containing self- and interaction dynamics, where $k_i^{\text{in}}$ is the in-degree of neuron $i$, and $\epsilon=1$. The second component represents a recovery variable where $a=0.28$, $b=0.5$ and $c=-0.04$.

The equations inferred by our approach from the neuronal activities data generated on a directed scale-free network are shown as Eq.\ (\ref{FHN-infer}). The activities and trajectories generated by the true Eq. (\ref{trueFHN}) and the inferred Eq. (\ref{FHN-infer}) are shown in Supplementary Fig.\ 5.
\begin{equation}
    \left\{
    \begin{aligned}
        \frac{d\hat{x}_{i,1}}{dt}&=0.989x_{i,1}-0.993x_{i,1}^3-0.996x_{i,2}-0.996\sum_{j=1}^{N}A_{ij}\frac{(x_{j,1}-x_{i,1})}{k_i^{\text{in}}},\\
        \frac{d\hat{x}_{i,2}}{dt}&=0.279+0.499x_{i,1}-0.040x_{i,2}.
    \end{aligned}
    \right.
    \label{FHN-infer}
\end{equation}
\begin{figure*}[!ht]
	\centering
	\includegraphics[width=0.65\textwidth]{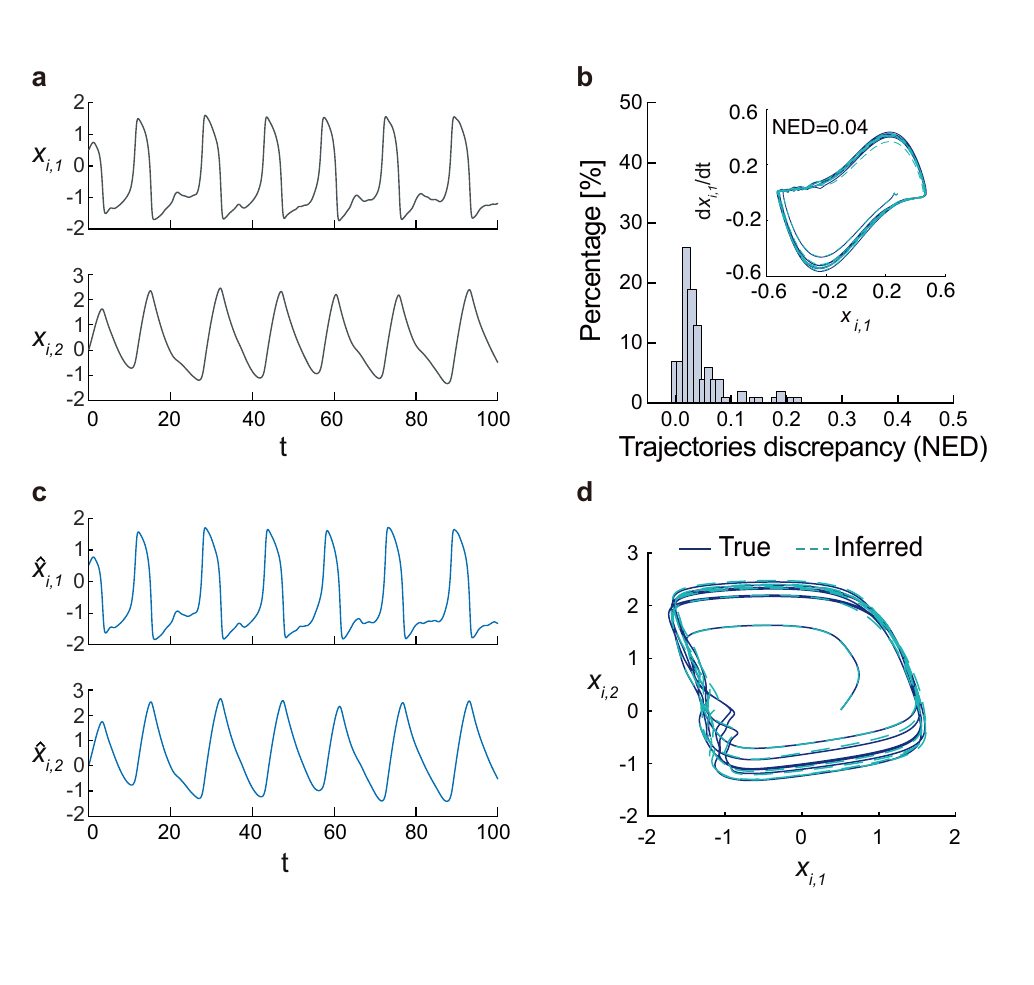}
	\caption{Neuronal activities and trajectories generated by the true FHN network dynamics and the inferred dynamical equations.}
	\label{FHNmodel}
\end{figure*}

\subsection{Gene regulation dynamics}
We also demonstrated the effectiveness of our approach in inferring gene regulation (GR) dynamics. We generate the gene activities data according to the equation\cite{mazur2009reconstructing}
\begin{equation}
    \frac{dx_{i,1}}{dt}=bx_{i,1}+\epsilon_{ij}\sum_{j=1}^NA_{i,j}\frac{x_{j,1}^{\alpha}}{x_{j,1}^{\alpha}+1},
\end{equation}
where $x_{i,1}(t)$ is the concentration of gene $i$ at time $t$, $\alpha=2$ denoting the Hill coefficient, and $b=-0.2$. Parameters $\epsilon_{ij}$ denotes the regulation strength of gene $j$ on gene $i$ which is set as $\epsilon_{ij}=0.1$ or $0.05$ on different networks. The dynamical interaction equation inferred by our approach from the data generated on an undirected scale-free network with $\epsilon_{ij}=0.05$ is
\begin{equation}
    \frac{d\hat{x}_{i,1}}{dt}=-0.197x_{i,1}+0.049\sum_{j=1}^NA_{i,j}\frac{x_{j,1}^{2}}{x_{j,1}^{2}+1}.
\end{equation}
The interaction function is automatically inferred from library $L_G$ that contains activation function $\{\frac{x^\gamma}{x^\gamma+1}\}$. The activities and trajectories generated by the true and the inferred equations are shown in Supplementary Fig.\ 6.
\begin{figure*}[!ht]
	\centering
	\includegraphics[width=0.7\textwidth]{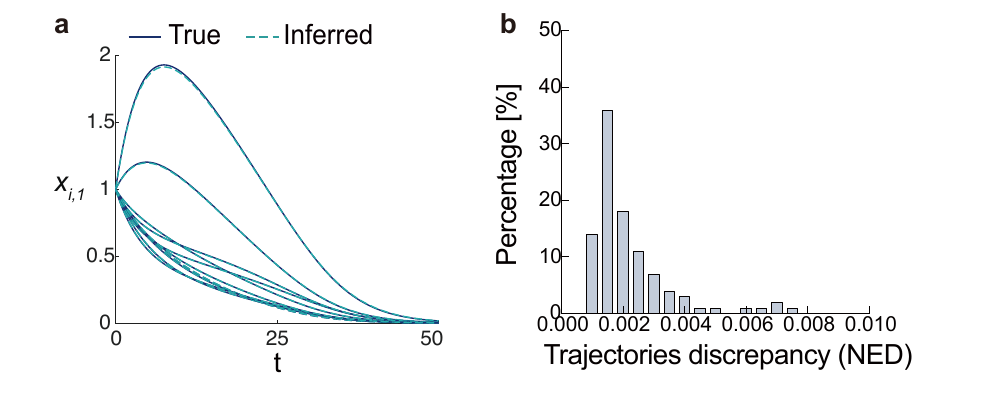}
	\caption{\textbf{a}. True and inferred activities for $10$ typical genes. \textbf{b}. The distribution of the discrepancy between true and inferred trajectories.}
	\label{generegulation}
\end{figure*}

\subsection{Coupled nonlinear oscillators}
We validated the two-phase inference approach for two types of coupled nonlinear oscillators, in which the self-dynamics are heterogeneous.
\subsubsection*{Coupled Kuramoto dynamics}
The first one is Kuramoto network dynamics\cite{schroder2017universal,pietras2019network}:
\begin{equation}
    \frac{dx_{i,1}}{dt}=\omega_i+\epsilon\sum_{j=1}^{N}A_{ij}\sin(x_{j,1}-x_{i,1})
\end{equation}
It describes the dynamics of $n$ oscillators with natural frequencies following a normal distribution $\omega\sim \mathcal{N}(1,\sigma)$ with $\sigma=0.1$, and the oscillators are coupled via the links $A_{ij}$. Parameter $\epsilon$ denotes the coupling strength of the interactions.

The equation inferred by our approach from the activities of oscillators generated on a directed scale-free network with $\epsilon=0.025$ is
\begin{equation}
    \label{kurainfer}
    \frac{d\hat{x}_{i,1}}{dt}=1.027+0.025\sum_{j=1}^{N}A_{ij}\sin(x_{j,1}-x_{i,1}),
\end{equation}
where constant $1.027$ is the inferred effective frequency $\hat{\omega}$ (\textit{i.e.} the average value $\langle \omega \rangle$ of nodes' natural frequencies). The activities and trajectories generated by the inferred equation are shown in Supplementary Fig.\ 7.
\begin{figure*}[!ht]
	\centering
	\includegraphics[width=0.75\textwidth]{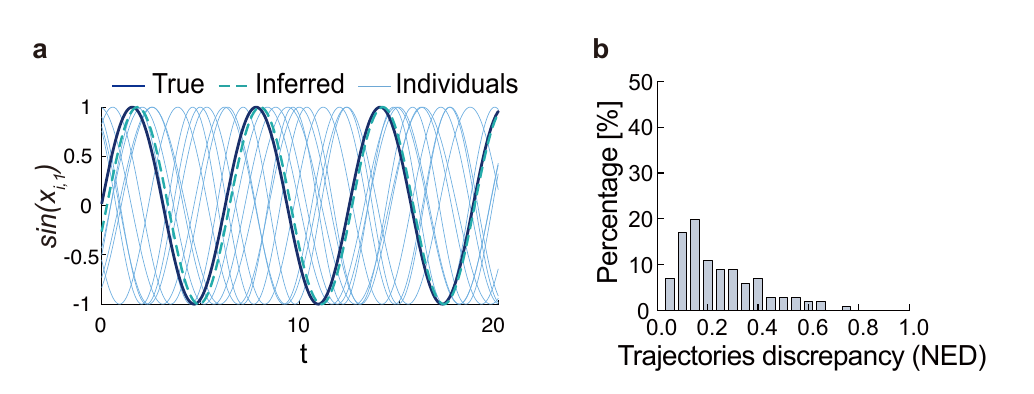}
	\caption{\textbf{a}. Activities denoted by $\sin{x_{i,1}}$, generated by the true and the inferred equations. Light blue curves show $10$ typical individuals' trajectories in the network, while the dark blue curve is the average of all nodes' trajectories. Dotted curve is the inferred activity. \textbf{b}. The distribution of the discrepancy between the true and inferred trajectories.}
	\label{Kuramoto}
\end{figure*}

\subsubsection*{Coupled R\"ossler dynamics}
Another case for coupled nonlinear oscillator system is R\"ossler network dynamics\cite{arenas2008synchronization, xiao2007phase, minati2019connectivity}. We first generated chaotic activities data according to the true equations governing \textit{heterogeneous} R\"ossler dynamics
\begin{equation}
    \left\{
    \begin{aligned}
    \frac{dx_{i,1}}{dt}&=-\omega_ix_{i,2}-x_{i,3}+\epsilon\sum_{j=1}^{n}A_{ij}(x_{j,1}-x_{i,1}), \\
    \frac{dx_{i,2}}{dt}&=\omega_ix_{i,1}+ax_{i,2}, \\
    \frac{dx_{i,3}}{dt}&=b+x_{i,3}(x_{i,1}+c),
    \end{aligned}
    \right.
\end{equation}
where coupling strength $\epsilon=0.15$ or $0.2$ on different networks, $a=0.2$, $b=0.2$, and $c=-6$. Oscillators' natural frequencies follow a normal distribution $\omega\sim \mathcal{N}(1,\sigma)$ with $\sigma=0.1$. The equations inferred by our approach from the data generated on a directed scale-free network with $\epsilon=0.15$ are
\begin{equation}
    \left\{
    \begin{aligned}
    \frac{d\hat{x}_{i,1}}{dt}&=-1.011x_{i,2}-1.009x_{i,3}+0.126\sum_{j=1}^{n}A_{ij}(x_{j,1}-x_{i,1}), \\
    \frac{d\hat{x}_{i,2}}{dt}&=1.019x_{i,1}+0.197x_{i,2}, \\
    \frac{d\hat{x}_{i,3}}{dt}&=0.196+0.993x_{i,3}x_{i,1}-5.653x_{i,3},
    \end{aligned}
    \right.
\end{equation}
where the coefficient $1.011$ is the inferred effective frequency $\hat{\omega}$ (\textit{i.e.} the average value $\langle \omega \rangle$ of nodes' natural frequencies). The activities and trajectories generated by the true and the inferred dynamical equations are shown in Supplementary Fig.\ 8.

\begin{figure*}[!ht]
	\centering
	\includegraphics[width=0.75\textwidth]{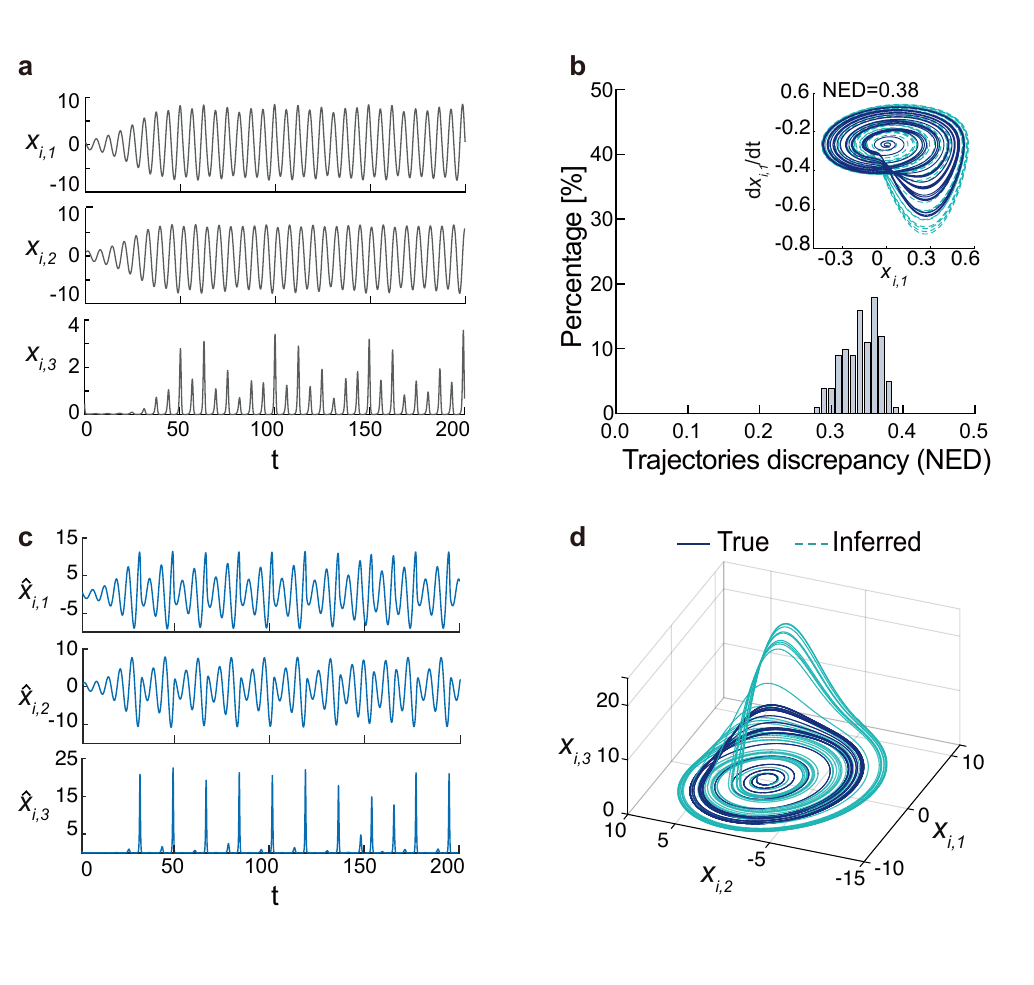}
	\caption{Oscillators' activities and trajectories generated by the true R\"ossler network dynamics and the inferred dynamical equations.}
	\label{Rosslerhetero}
\end{figure*}

We also generated \textit{homogeneous} R\"ossler dynamics according to the true equations
\begin{equation}
    \left\{
    \begin{aligned}
    \frac{dx_{i,1}}{dt}&=-x_{i,2}-x_{i,3}+\epsilon\sum_{j=1}^{n}A_{ij}(x_{j,1}-x_{i,1}), \\
    \frac{dx_{i,2}}{dt}&=x_{i,1}+ax_{i,2}, \\
    \frac{dx_{i,3}}{dt}&=b+x_{i,3}(x_{i,1}+c),
    \end{aligned}
    \right.
\end{equation}
where $\epsilon=0.1$, $a=0.35$, $b=0.2$ and $c=-5.7$. The equations inferred by our approach from the data generated on a directed scale-free network are
\begin{equation}
    \left\{
    \begin{aligned}
    \frac{dx_{i,1}}{dt}&=-0.999x_{i,2}-0.999x_{i,3}+0.100\sum_{j=1}^{n}A_{ij}(x_{j,1}-x_{i,1}), \\
    \frac{dx_{i,2}}{dt}&=0.999x_{i,1}+0.349x_{i,2}, \\
    \frac{dx_{i,3}}{dt}&=0.181+x_{i,3}(0.993x_{i,1}-5.645),
    \end{aligned}
    \right.
\end{equation}

\begin{figure*}[!ht]
	\centering
	\includegraphics[width=0.9\textwidth]{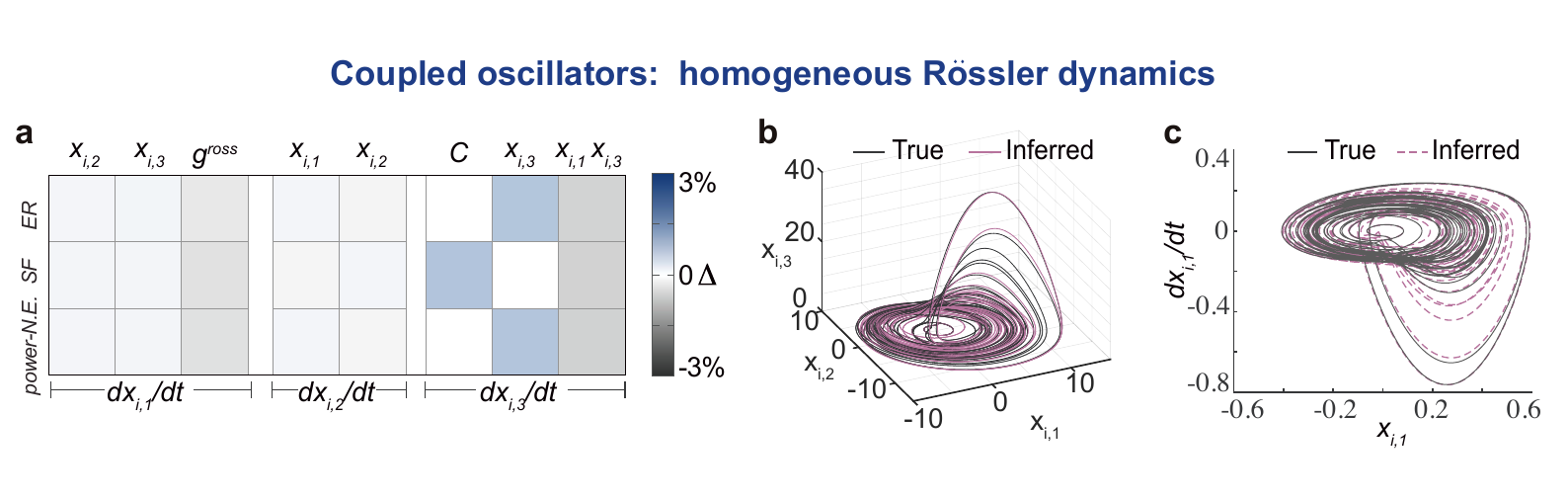}
	\caption{\textbf{a}. Relative errors $\Delta$ of the inferred elementary functions and their coefficients. \textbf{b,c}. Nodes' activities and trajectories generated by the true and the inferred dynamical equations.}
	\label{Rosslerhomo}
\end{figure*}
To illustrate the generative power of the inferred equations obtained by our approach, we set different initial states to validate if the inferred equations can reproduce the dynamics of the original model. The results in Supplementary Fig.\ 10 show that the inferred equations indeed resemble the dynamics of original models. It is worth mentioning that although our approach infers the true equations with high precision, the distance between the inferred and the true trajectories increases for longer time due to the chaoticity of the original model.
\begin{figure*}[!ht]
	\centering
	\includegraphics[width=0.8\textwidth]{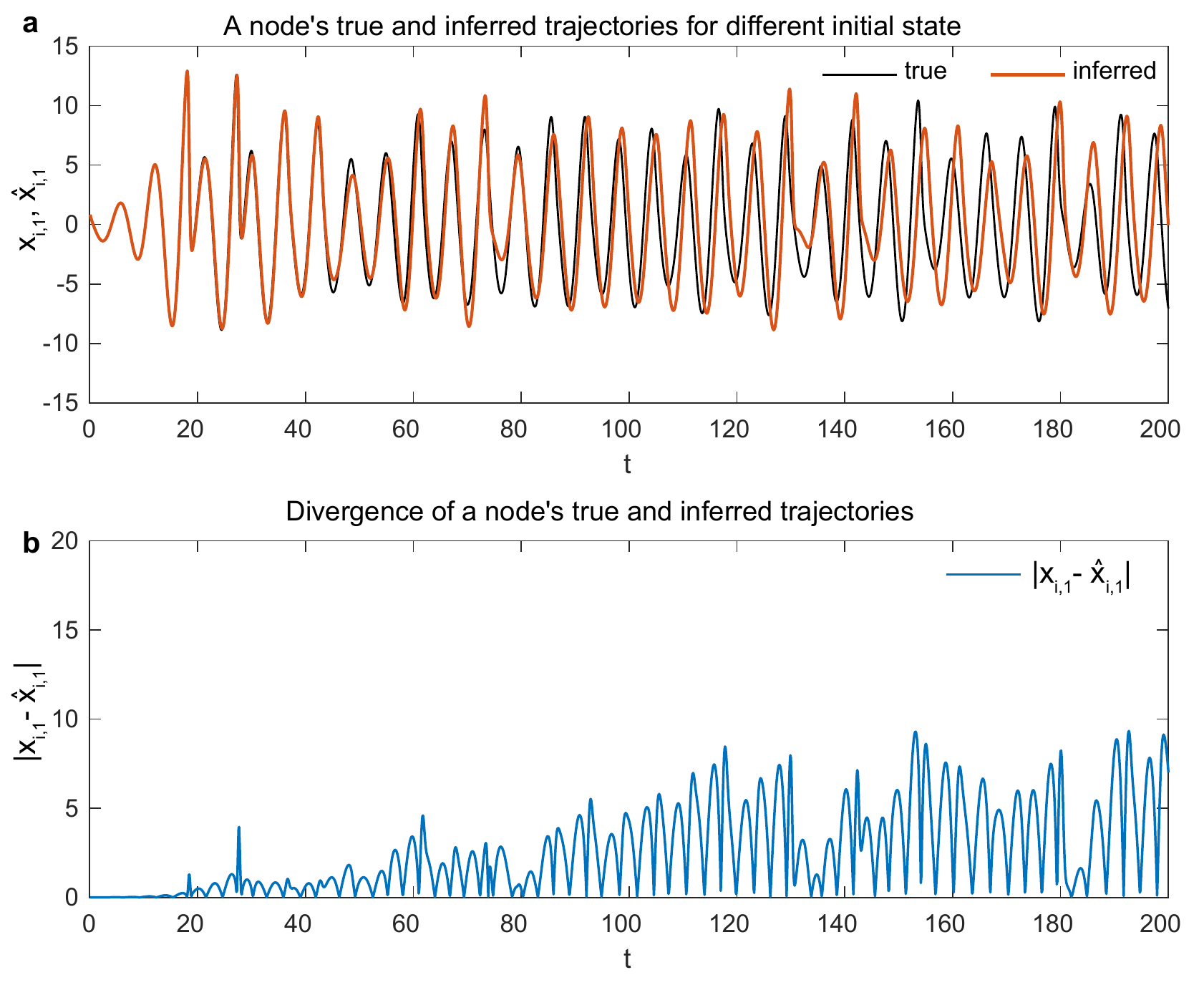}
	\caption{\textbf{a}. A node's $x$-dimension trajectories generated by the true and the inferred dynamical equations for an initial state that is different from the initial state of original data. \textbf{b}. Divergence of trajectories generated by the true and the inferred dynamical equations.}
	\label{Rosslerdiv}
\end{figure*}


\clearpage
\subsection{Edge dynamics of social balance}
We also tested our approach with a social balance dynamics described by edge dynamical equation\cite{marvel2011continuous}, \textit{i.e.} the relation between two individual nodes is represented by $x_{ij}$. Entry $x_{ij}$ represents the strength of trust or distrust between $i$ and $j$.
The strength of trust or distrust between nodes evolves according to
\begin{equation}
    \Dot{\mathbf{X}}=\mathbf{X}^2.
    \label{SB dynamics}
\end{equation}
We rewrite Eq. (\ref{SB dynamics}) explicitly in terms of entries $x_{ij}$ as
\begin{equation}
    \frac{d{x_{ij}}}{d{t}}=\sum_{k}x_{ik}x_{kj}.
\end{equation}
The edge activities generated by the true equation are shown in Supplementary Fig.\ 11. To infer such a equation, we construct libraries by a variety of functions with temporal topologies data $\mathbf{X}(t)$, for example, $\mathbf{X}^2$, $\mathbf{X}^3$, $\sin(\mathbf{X})$, \textit{etc}. By our approach, we automatically infer the exact term and its coefficient, leading to the equation
\begin{equation}
    \label{inferSB}
    \frac{d{\hat{x}_{ij}}}{d{t}}=0.999\sum_{k}x_{ik}x_{kj}.
\end{equation}
\begin{figure*}[!ht]
	\centering
	\includegraphics[width=0.35\textwidth]{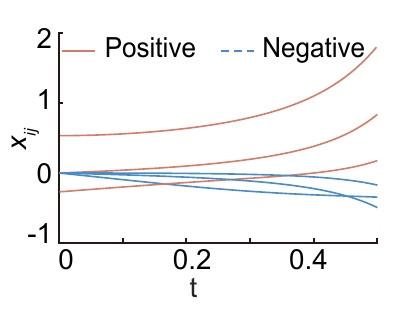}
	\caption{Activities of six typical edges generated by the inferred Eq. (\ref{inferSB}).}
	\label{SB}
\end{figure*}
\clearpage
\subsection{Reproducibility description}
To ensure the reproducibility of our results we list the parameters of dynamics and networks for each plot in the main paper. The simulations of dynamics are from time $0$ to $T$ with step-size $\delta t$.
\begin{table}[!ht]
  \begin{center}
    \caption{Parameters setting in this paper.}
    \label{Parameters Table}
    \resizebox{.7\columnwidth}{!}{
    \begin{tabular}{l|c|c|c}
      \toprule
      \hline
      Plots&Dynamics&Networks&Simulation data\\
      \hline
      Fig.\ 1\textbf{b-f} &HR, $\epsilon_1^{HR}=0.30$, $\epsilon_2^{HR}=-0.15$&BA ($\gamma=3$)&\makecell{$T=500$\\$\delta t=0.01$}\\
      \hline
      Fig.\ 2 &FHN, $\epsilon^{FHN}=1.00$&ER, SF($\gamma=2.5$),\textit{Drosophila}&\makecell{$T=140$\\$\delta t=0.01$}\\
      \hline
      Fig.\ 3\textbf{a}&HR, $\epsilon_1^{HR}=0.30$, $\epsilon_2^{HR}=-0.15$&ER, SF ($\gamma=2.5$),\textit{C.elegans}&\makecell{$T=500$\\$\delta t=0.01$}\\
      \hline
      Fig.\ 3\textbf{c}&S.B., $\epsilon^{SB}=1.00$&ER, SF($\gamma=2.5$),Advogato&\makecell{$T=0.7$\\$\delta t=0.001$}\\
      \hline
      Fig.\ 3\textbf{e}&\makecell{Kuramoto, $\epsilon^{kura}=0.015$\\$\sigma=0.1$}&ER&\makecell{$T=100$\\$\delta t=0.01$}\\
        \hline
      Fig.\ 3\textbf{e}&\makecell{Kuramoto, $\epsilon^{kura}=0.03$\\$\sigma=0.1$}&SF ($\gamma=2.5$),U.S. power gird&\makecell{$T=100$\\$\delta t=0.01$}\\
      \hline
      Fig.\ 3\textbf{g}&\makecell{R\"ossler, $\epsilon^{ross}=0.15$\\$\sigma=0.1$}&ER, SF ($\gamma=2.5$)&\makecell{$T=100$\\$\delta t=0.01$}\\
      \hline
      Fig.\ 3\textbf{g}&\makecell{R\"ossler, $\epsilon^{ross}=0.10$\\$\sigma=0.1$}&N.E. power grid&\makecell{$T=100$\\$\delta t=0.01$}\\
      \hline
      Fig.\ 4\textbf{a}&FHN, $\epsilon^{FHN}=(0.5,20)$&BA ($\gamma=3$)&\makecell{$T=100$\\$\delta t=0.01$}\\
      \hline
      Fig.\ 4\textbf{b}&\makecell{Kuramoto, $\epsilon^{kura}=0.03$\\$\sigma=(0.1,1)$}&BA ($\gamma=3$)&\makecell{$T=100$\\$\delta t=0.01$}\\
      \hline
      Fig.\ 4\textbf{c}&HR, $\epsilon_1^{HR}=0.30$, $\epsilon_2^{HR}=-0.15$&SF ($\gamma=2.5$)&\makecell{$T=500$\\$\delta t=0.01$}\\
      \hline
      Fig.\ 4\textbf{c}&GR, $\epsilon^{GR}=0.05$&BA ($\gamma=3$)&\makecell{$T=50$\\$\delta t=0.01$}\\
      \hline
      Fig.\ 5\textbf{a-j}&HR, $\epsilon_1^{HR}=0.30$, $\epsilon_2^{HR}=-0.15$&BA ($\gamma=3$)&\makecell{$T=500$\\$\delta t=0.01$}\\
      \hline
      Fig.\ 5\textbf{k}&\makecell{HR, $\epsilon_1^{HR}=0.30$, $\epsilon_2^{HR}=-0.15$\\observational $60$ dB, missing $40\%$}&BA ($\gamma=3$)&\makecell{$T=500$\\$\delta t=0.01$}\\
      \hline
      Fig.\ 5\textbf{l}&\makecell{R\"ossler, $\epsilon^{ross}=0.20$\\$\sigma=0.1$\\observational $40$ dB, spurious $20\%$}&BA ($\gamma=3$)&\makecell{$T=100$\\$\delta t=0.01$}\\
      \hline
      \bottomrule
    \end{tabular}}
  \end{center}
\end{table}

In Fig.\ 5 of the main paper $\sigma=0.1$ for Kuramoto and R\"ossler network dynamics, and the simulations of dynamics for each model are same as those in Figs.\ 2-3.

\section{Details of inferrability}
\subsection{Quantifying synchronization}
To quantify the degree of synchronization of a dynamical network, the sequence of peaks in $x_i(t)$ is extracted for defining the geometric phase $\theta_i(t)$ of each node $i$. For example, in the neuronal spiking activities data, we use a Poincar\'{e}'s section at $x_i=0.5$ to determine the time of the start (upward sense) and the end (downward sense) of a spike (Supplementary Fig.\ 12(a,c)). Between the $l$-th spike and $(l+1)$-th spike, the phase $\theta_i(t)$ is increased by $2\pi$, such that an interpolation between the two spikes can define the continuous time-varying phase as\cite{boaretto2019mechanism,checco2008synchronization}
\begin{equation}
    \theta_i(t)=2\pi l_i+2\pi\frac{t-t_{l,i}}{t_{l+1,i}-t_{l,i}}, \hspace*{0.5cm} t_{l,i}\leq t<t_{l+1,i},
\end{equation}
where $t$ is the current time, and $t_{l,i}$ is the time at which node $i$ starts the $l$-th spike. The degree of synchronization of the dynamical network is then quantified by order parameter
\begin{equation}
    R(t)=\left|\frac{1}{N}\sum_{i=1}^Ne^{i\theta_i(t)}\right|.
\end{equation}

When $R(t)$ is close to zero, the dynamical network is totally disorder; If $R(t)=1$, the activities of all nodes are completely synchronized, indicating a complete phase synchronized state. Averaging $R(t)$ over time we obtain a single order parameter to quantify network synchronization, \textit{i.e.}
\begin{equation}
    \langle R \rangle=\frac{1}{t_f-t_0}\sum_{t=t_0}^{t_f}R(t),
\end{equation}

where $t_0$ and $t_f$ are the start and the end of the time range respectively. A network usually becomes more synchronized when the coupling strength increases. It is worth noting that, for some network dynamics, such as FHN, the nodes can evolve into two categories, \textit{i.e.} get trapped in two fixed points (Supplementary Fig.\ 12e). For this case, the order parameters of the two categories are calculated respectively and the average of them is considered as the order parameter of the whole network.

\begin{figure}[!ht]
\centering
\includegraphics[width=0.9\linewidth]{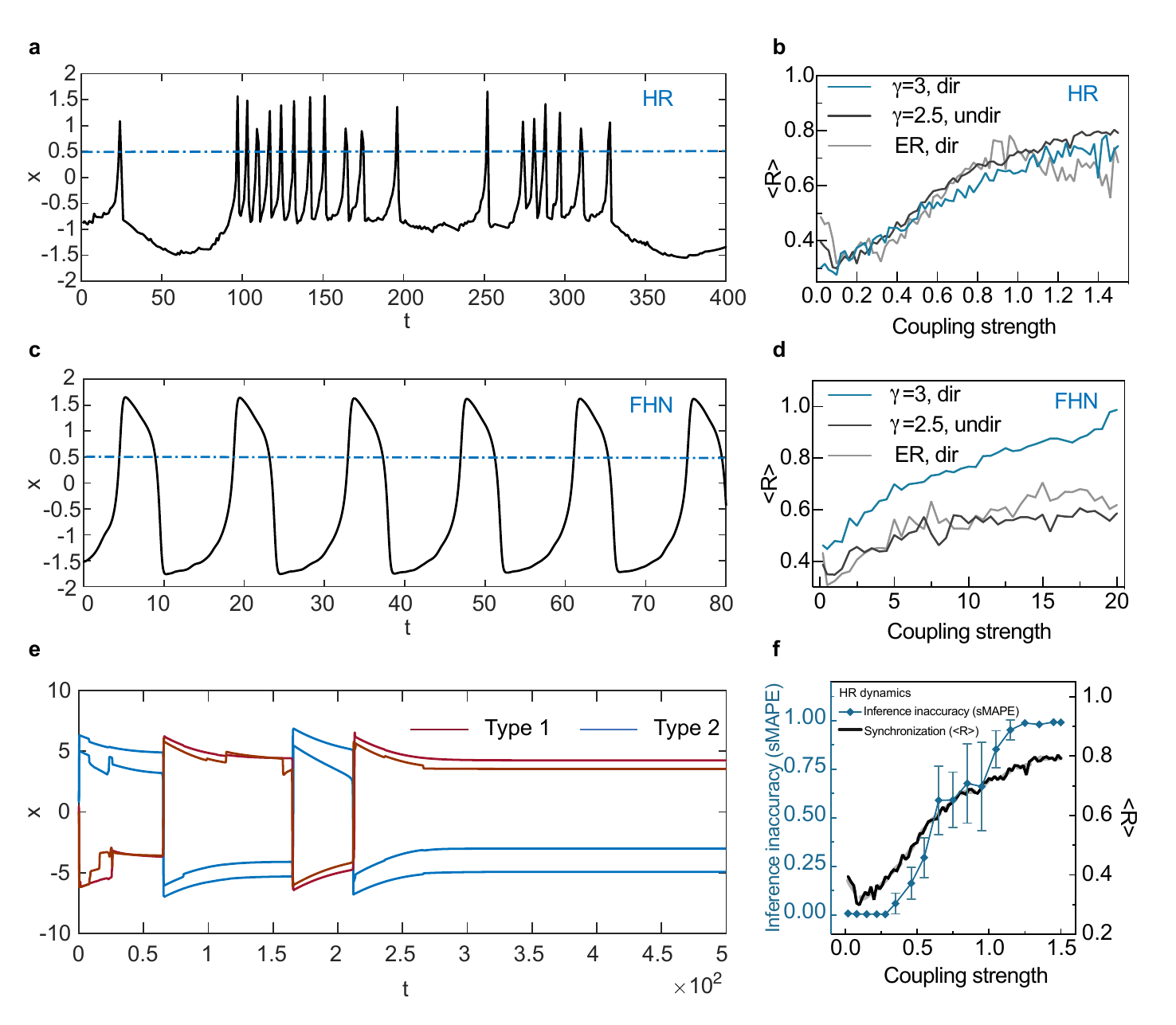}
\label{synchronization}
\caption{Quantification of synchronization for spiking activities. \textbf{a}. Poincar\'e's section of HR neuronal dynamics segmented by the blue dotted line. \textbf{b}. Order parameter $\langle R \rangle$ of HR network dynamics vs. coupling strength for different network. \textbf{c}. Poincar\'e's section of FHN neuronal dynamics segmented by the blue dotted line. \textbf{d}. Order parameter $\langle R \rangle$ of FHN network dynamics vs. coupling strength. \textbf{e}. Two categories of $\boldsymbol{x}(t)$, obtanined by clustering. \textbf{f}. Inference inaccuracy (sMAPE) increases when the coupling strength becomes larger, indicating that more synchronized is a network, more difficult to infer its hidden interaction mechanisms.}
\centering
\end{figure}

\subsection{Inferring alternative equations}
If the hidden true dynamics is not a combination of elementary functions or if some elementary function of the hidden dynamics is not included in the large libraries $L_F$ and $L_G$, our approach will infer alternative forms of $\textbf{F}$ and $\textbf{G}$ that are still able to characterize the observed system behaviors. To demonstrate such capacity, we deliberately remove some elementary functions that should exist in the true network dynamics. For example, we delete interaction terms $\frac{x_{i,1}}{1+e^{-\alpha(x_{j,1}-\beta)}}$ and $\frac{1}{1+e^{-\alpha(x_{j,1}-\beta)}}$ ($\alpha=10$, $\beta=1$), yet to infer HR network dynamics with $r=0.004$ and $x_0=-1.5$. Even for such deficient libraries our approach infers alternative equations
\begin{equation}
    \left\{
    \begin{aligned}
    \frac{d\hat{x}_{i,1}}{dt}&=2.903x_{i,1}^2-1.006x_{i,1}^3+0.981x_{i,2}-0.985x_{i,3}-0.043\sin(x_{i,1})+3.211\\
    &+0.119\sum_{j=1}^NA_{ij}\frac{1}{1+e^{-x_{j,1}}}+0.053\sum_{j=1}^NA_{ij}\sin(x_{j,1}),\\
    \frac{d\hat{x}_{i,2}}{dt}&=1.000-4.994x_{i,1}^2-0.998x_{i,2}, \\
    \frac{d\hat{x}_{i,3}}{dt}&=0.024+0.016x_{i,1}-0.004x_{i,3},
    \end{aligned}
    \right.
\end{equation}
that still capture the HR network dynamics (see Fig.\ 4c in the main paper), as well as the alternative equation
\begin{equation}
    \begin{split}
        \frac{d\hat{x}_{i,1}}{dt}&=-0.0185-0.11\sin{x_{i,1}}-0.09x_{i,1}^2+0.01\sum_{j=1}^NA_{ij}\frac{x_{j,1}x_{i,1}}{x_{j,1}x_{i,1}+1}\\
        &+0.012\sum_{j=1}^NA_{ij}\frac{x_{j,1}^5}{x_{j,1}^5+1}+0.017\sum_{j=1}^NA_{ij}\sin{x_{j,1}}
    \end{split}
\end{equation}
to capture GR network dynamics (see also Fig.\ 4c).

\subsection{Inferrability comparison studies}
In this subsection we compare our approach with SINDy\cite{brunton2016discovering} and its variant\cite{mangan2017model} regarding inferrability, \textit{i.e.}\ the capacity of dealing with partially synchronized networked dynamics or dynamical heterogeneity. The results showing in Supplementary Fig.\ 13 indicate that our method is better at coping with synchronization and dynamical heterogeneity. The network properties and dynamics settings are the same as that in Fig.\ 4 of the main paper.
\begin{figure*}[!ht]
	\centering
	\includegraphics[width=0.65\textwidth]{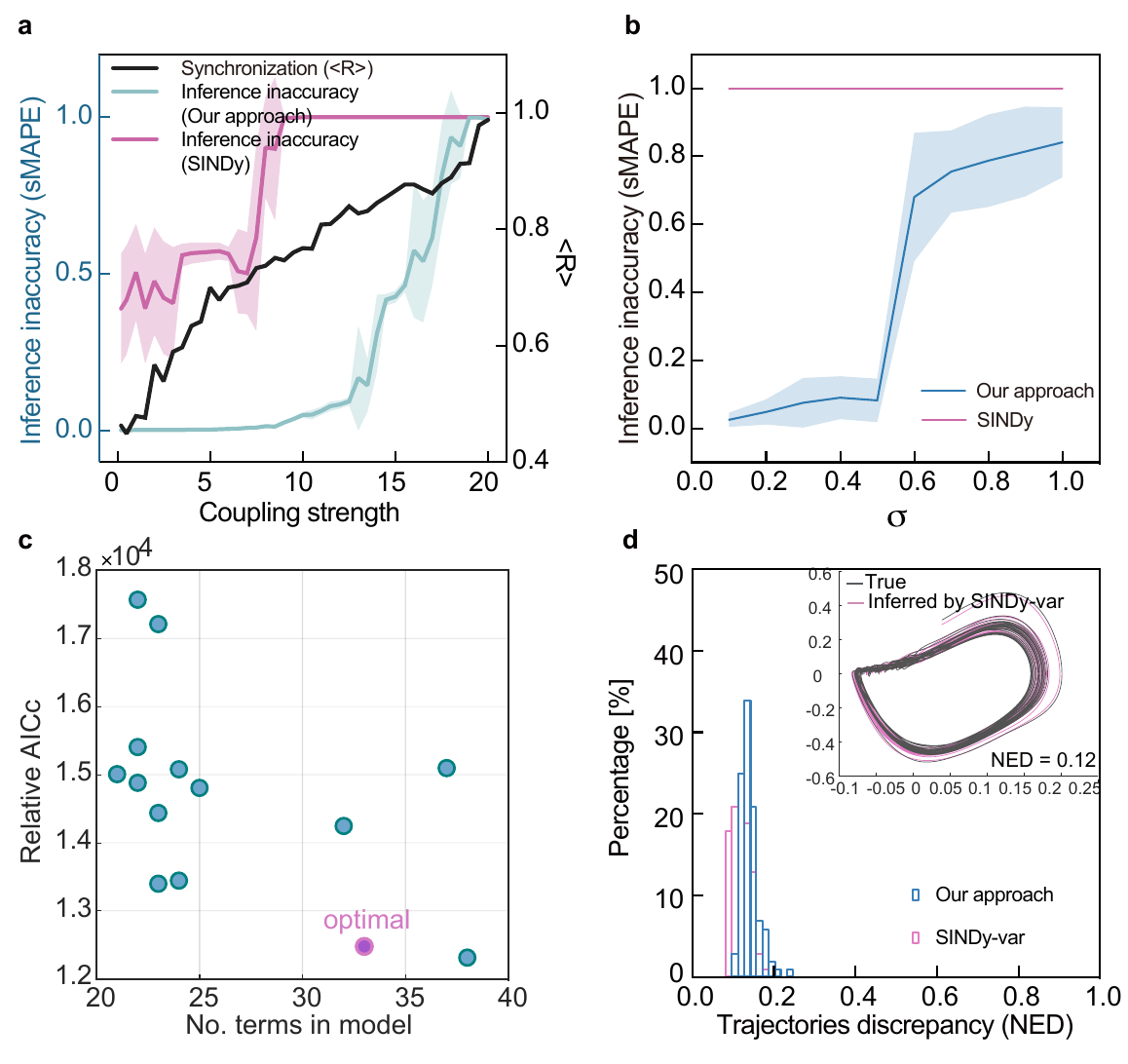}
	\caption{Inferrability comparison between SINDy, its variant and our approach regarding synchronization (FHN dynamics; \textbf{a}), dynamical heterogeneity (Kuramoto dynamics; \textbf{b}), and deficient libraries (HR dynamics; \textbf{c,d}). \textbf{c}. The alternative equations obtained by SINDy-variant, which selects the smallest number of terms and lowest relative AIC value, resulting in an equation containing $34$ terms. \textbf{d}. Comparison of discrepancy between SINDy-variant and our approach. Note that the number of terms in the alternative equation inferred by our approach is $14$, much smaller than $34$ by SINDy-variant yet similar to $13$ in the true equation.}

\end{figure*}

\section{Details of robustness analyses}
In this work we validated the robustness of our approach against five types of unavoidable uncertainties in observation data, including observational, uncorrelated and correlated dynamical noises in data of nodes' activities, missing and spurious links in observed topologies, low resolution by experimental techniques. To construct incomplete topologies we randomly add or delete a proportion of entries in the true adjacency matrix $A_{ij}$; To imitate the constraint of low resolution we regularly down-sample the time series of nodes' activities. In the following we describe in detail the simulations of data with observational, dynamical and correlated noise, and show the failure ratio of inferring the structure of true elementary functions out of $100$ independent runs in Supplementary Fig.\ 14.

\subsection{Noisy data}
There are usually two types of noises in time series data, \textit{i.e.}\ dynamical noise and observational noise\cite{sase2016estimating}. The latter is raised in the observation or measurement process, while the former indicates the inherent stochasticity of system dynamics. To test the robustness of our approach against these two types of noises, we simulate the data through\cite{sase2016estimating}
\begin{equation}
\label{noise}
    \left\{
    \begin{aligned}
    \frac{d\boldsymbol{x}_i(t)}{dt}&=F(\boldsymbol{x}_i(t))+\sum_{j=1}^nA_{ij}G(\boldsymbol{x}_i(t),\boldsymbol{x}_j(t))+\eta A\cdot dW_i(t), \\
    X^{obs}(t)&= \boldsymbol{x}_i(t)+a\beta_X(t),
    \end{aligned}
    \right.
\end{equation}
where $\eta A\cdot dW_i(t)$ is dynamical noise, and $a\beta_X(t)$ is observational noise. Here $dW_i(t)$ is the Gaussian white noise, which follows a normal distribution with mean zero and standard deviation $\sqrt{dt}$, and $A$ is the average amplitude of the original time series. Here $\eta$ represents the intensity of dynamical noises relative to the signal amplitude. We simulate Eq. (\ref{noise}) by using the fourth-order Runge-Kutta method with fixed time step size. The $\beta_X(t)$ is a Gaussian noise following a normal distribution with zero mean and standard deviation one, hence the intensity of observational noise can be tuned by the value of $a$. In the paper we generate observational noisy data by awgn in \textit{MatLab} and characterize the noisy data with specified Signal-to-Noise-Ratio (SNR).

Correlated dynamical noise could exist in complex systems when the noises are correlated across different nodes (such as in neuronal activities\cite{eyherabide2013and,van2018modeling}). Then the network dynamics become
\begin{equation}
    \frac{d\boldsymbol{x}_i(t)}{dt}=F(\boldsymbol{x}_i(t))+\sum_{j=1}^nA_{ij}G(\boldsymbol{x}_i(t),\boldsymbol{x}_j(t))+\epsilon_i,
\end{equation}
where $\boldsymbol{\epsilon}=\{\epsilon_{i}\}$ is correlated noise across all nodes drawn from a multivariate normal distribution with mean $0$ and covariance $\boldsymbol{\Sigma}$, \textit{i.e.}
\begin{equation}
    \boldsymbol{\epsilon}\sim\eta'\mathcal{N}(\textbf{0},\boldsymbol{\Sigma}).
\end{equation}
The noise covariance matrix $\boldsymbol{\Sigma}$ is semidefinite which is constructed by $\boldsymbol{\Sigma}=AA^{\top}$, where $A$ is adjacency matrix, and $\eta'$ is the strength of correlated noise. In the main paper we also validate the robustness of our approach against such correlated noises (Fig.\ 5j).

\begin{figure*}[!ht]
	\centering
	\includegraphics[width=0.9\textwidth]{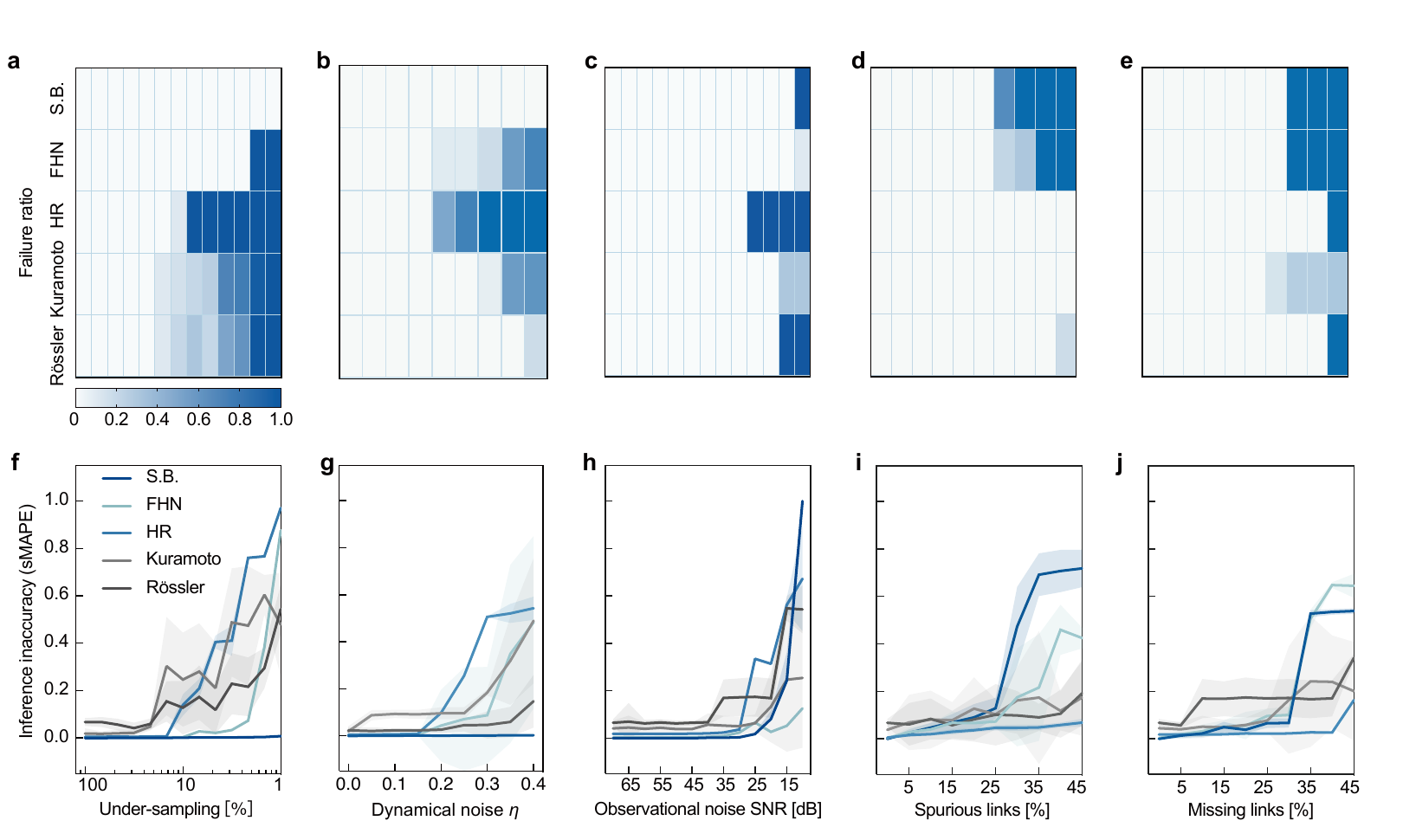}
	\caption{The failure ratio of inferring the structure of true dynamics of $100$ independent runs (\textbf{a-e}), and the inference inaccuracies sMAPE of inferring both true elementary functions and their coefficients (\textbf{f-j}).)}

\end{figure*}

\subsection{Ablation studies}
In the main paper we showed that each key sub-step of our two-phase approach is indispensable for inferring the hidden network dynamics from noisy and incomplete data. Here we further show that, even for clean data these key sub-steps are also necessary. To do so we individually remove each sub-step and test the performance of the corresponding deficient approach. The result of these ablation studies are shown in Supplementary Fig.\ 15.

\begin{figure}[!ht]
\centering
\includegraphics[width=\linewidth]{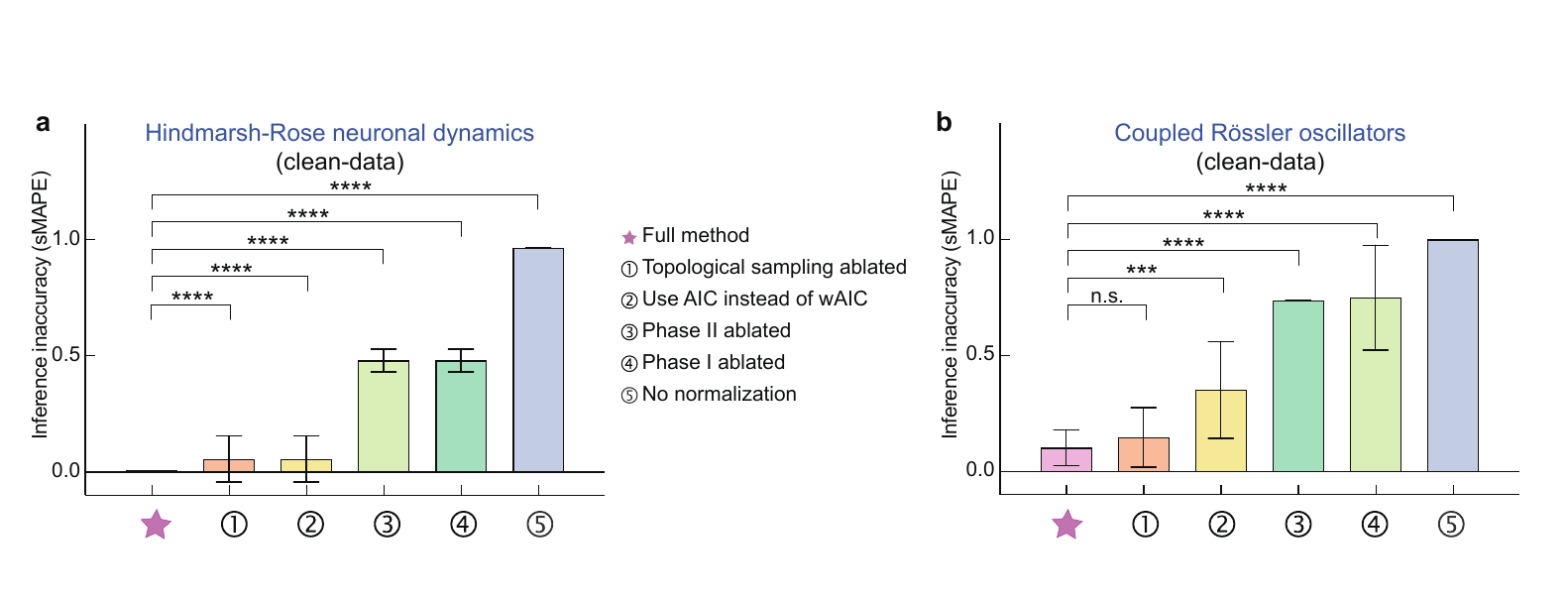}
\label{clean-ablated}
\caption{The coupled results of ablation studies on clean data, for inferring the hidden equations governing HR network dynamics (\textbf{a}) and coupled R\"ossler dynamics (\textbf{b}). Five ablation studies were performed and shown in box-whisker plot: \textcircled{1} removing the sub-step of topological sampling, \textcircled{2} using original AIC instead of wAIC, \textcircled{3} removing Phase \uppercase\expandafter{\romannumeral2}, \textcircled{4} removing Phase \uppercase\expandafter{\romannumeral1}, \textcircled{5} No normalization in $\Theta_F$ and $\Theta_G$. Statistical significance is obtained by multiple Mann-Whitney test\cite{cheung1997mann,klotz2006computational}. Three or four asterisks indicate $p$-value $10^{-3}$ or $<10^{-4}$, and n.s. means not significant.}
\centering
\end{figure}

\textcircled{1} Topological sampling ablated: In Phase \uppercase\expandafter{\romannumeral2} we randomly select sets of single node and its neighbors, instead of the all nodes. Such topological sampling can imitate the incompleteness in observed topology and increase the accuracy of the approach. If this sub-step is removed, the accuracy indeed decreases significantly.

\textcircled{2} Using AIC instead of wAIC: As shown in Supplementary Table \ref{phase2 importance evidence}, AIC could rank the relevance of elementary functions incorrectly. Indeed, it brings also several irrelevant terms. In contrast, we find that wAIC can successfully identify the most relevant terms and, importantly, there is a clear threshold seperating the relevant and irrelevant terms. Hence, using AIC instead of wAIC causes significantly increased inaccuracy.

\textcircled{3} Phase \uppercase\expandafter{\romannumeral2} ablated:
Without the local fine-tuning, Phase \uppercase\expandafter{\romannumeral1} alone is not able to obtain a compact form of the hidden equation. Moreover, as shown in Fig.\ 1d in the main paper, even though the equation inferred by Phase \uppercase\expandafter{\romannumeral1} alone fits the observational data well, it does not have generative power.

\textcircled{4} Phase \uppercase\expandafter{\romannumeral1} ablated:
If Phase \uppercase\expandafter{\romannumeral1} is ablated from the approach, we will lose the global information that captures the coarse yet consistent structure of the hidden dynamics. Hence we find the ablation of Phase \uppercase\expandafter{\romannumeral1} also significantly increase the inference inaccuracy.

\textcircled{5} No normalization in Phase \uppercase\expandafter{\romannumeral1}:
As shown in Supplementary Fig.\ 3, the value of several elementary functions can span several orders of magnitude, possibly significantly larger than that of others. If the sub-step of normalization is removed the coefficients of these inherently high-value terms, obtained by regression will be very small, which decreases the possibility of the existence of these terms.

 \begin{table}[!ht]
  \begin{center}
  \caption{The example showing the results of using AIC vs. wAIC for automatically choosing the relevant elementary functions. Ten leading elementary functions are ranked by their absolute coefficients. Blue color denotes those which should be selected to compose the true governing equation.}
    \label{phase2 importance evidence}
    \resizebox{.75\columnwidth}{!}{
    \begin{tabular}{c|c|c|c}
      \toprule
      \hline
      elementary functions&coefficients&AIC[$\times 10^6$]&wAIC[$\times 10^7$]\\
      \hline
      \textcolor[RGB]{56,108,176}{$x_{i,2}$}&-1.0&0.28&0.29\\
      \textcolor[RGB]{56,108,176}{$x_{i,3}$}&-0.8&-0.16&-0.002\\
      $\sin(x_{i,3})$&-0.08&-0.16&-0.02\\
      \textcolor[RGB]{56,108,176}{$x_{j,1}-x_{i,1}$}&0.07&-0.13&-0.01\\
      $x_{i,1}$&-0.03&-0.15&-0.05\\
      $x_{i,1}x_{i,3}$&-0.02&-0.04&-0.02\\
      $\frac{1}{1+e^{-10[(x_{j,1}-x_{i,1})-1]}}$&0.01&-0.16&-0.3\\
      $\frac{1}{1+e^{-5[(x_{j,1}-x_{i,1})-1]}}$&0.00&-0.16&-1.4\\
      $e^{x_{j,1}-x_{i,1}}$&0.00&-0.16&-166\\
      $e^{x_{j,1}}$&0.00&-0.20&-1750\\
      \hline
      \bottomrule
    \end{tabular}}
  \end{center}
\end{table}

\subsection{Robustness comparison studies}
In the main paper we show the results of the comparisons between our approach, SINDy, and ARNI. The original ARNI is a model-free inference approach for network structure reconstruction\cite{casadiego2017modelSI}. ARNI uses the idea of matrix pseudoinverse to connect a node to other most relevant nodes by minimizing the cost between numerical derivatives of observed time series and the sum value of basis functions, enabling the inference of network topology from nodes' activities. We would like to emphasize that although ARNI originally aimed at inferring network structure, it can also be used to infer network dynamics if network structure is given. For a fair comparison, we transfer ARNI to infer network dynamics as follows. We apply the same idea of matrix pseudoinverse and add most relevant elementary functions one by one, as shown in Supplementary Fig.\ 16, until the cost is lower than a threshold. To improve ARNI's ability of coping with the possible incompleteness in observed network structure, we also perform topological samplings in ARNI. The resulting elementary functions and their coefficients are the value averaged for multiple topological samplings. The details of ARNI modification can be seen in our code.
\begin{figure*}[!ht]
	\centering
	\includegraphics[width=0.6\textwidth]{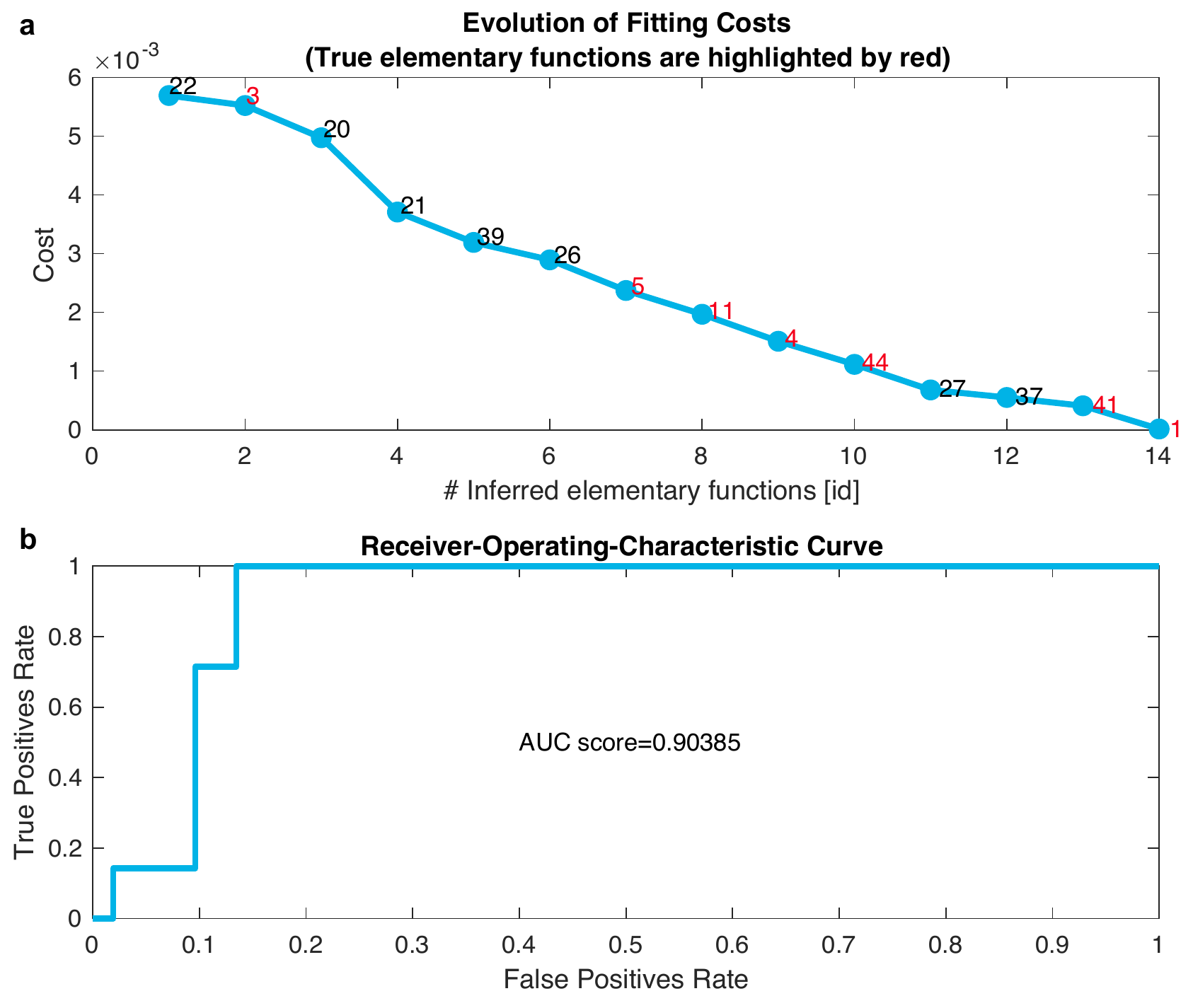}
	\caption{Inference of elementary functions by ARNI. \textbf{a}. The cost curve when most relevant elementary functions are added one by one. The IDs of the true elementary functions are highlighted by red. \textbf{b}. The accuracy of the inferred elementary functions is evaluated by AUC score.}
\end{figure*}

We also perform robustness comparisons between our approach and SINDy variant\cite{mangan2017modelSI} against observational noises. The results are shown in Supplementary Fig.\ 17, where the task is to inferring HR dynamics from simulated activity data with 40dB observational noise. The simulation time $T=500$ with step size $\delta t=0.01$. The network is generated with directed BA model with size $n=100$ and average degree $\langle k\rangle=5$.
\begin{figure*}[!ht]
	\centering
	\includegraphics[width=0.8\textwidth]{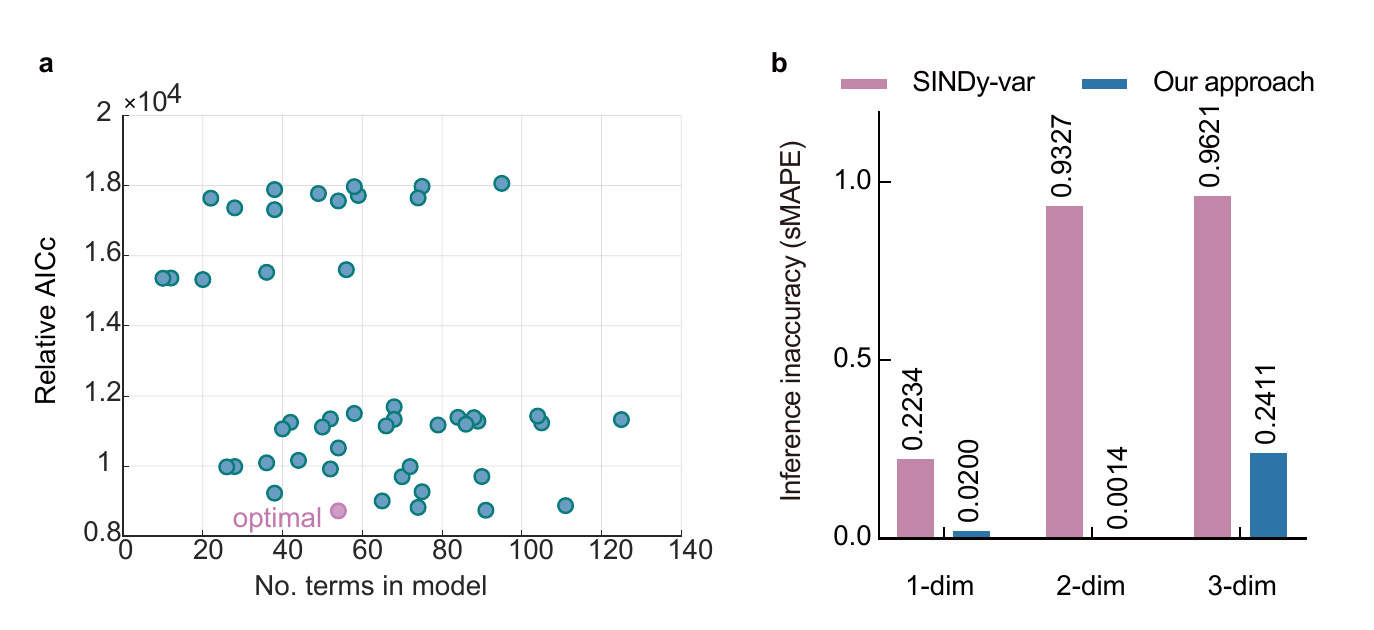}
	\caption{Inferring HR dynamics from simulated activity data with $40$dB observational noise by SINDy variant\cite{mangan2017model} and our approach respectively. \textbf{a}. Number of terms in inferred model by SINDy variant vs. the relative AIC. SINDy variant selects the optimal model with the lowest relative AIC value and the smallest number of terms (highlighted by purple). \textbf{b}. Inference inaccuracies comparison between SINDy variant and our approach.}

\end{figure*}
\clearpage
\section{Details of empirical system inference}
In the main text we have demonstrated the applicability of our approach to inferring dynamical equations from empirical data. Specifically we used the global spreading data of H1N1 disease reported daily from April 24th to July 6th in year $2009$. As we focused on the dynamics of early spreads before governments introduce quarantine policies, only the first $45$ daily data were used for the inference. For instance, if the first case in a country was reported on May 1st, we employed the data from May 1st to June 14th. Note that, although the starting time of spread for different countries are different, in the coupling dynamics $G(x_i,x_j)$ the time $t$ is the same date for all nodes. In other words, in Fig.\ 6 and Supplementary Figs. 18-20 the initial time $t$ [/day] being shifted to $t=1$ for all nodes are just for the convenience of visualization. The inference approach indeed used the original dates and data. The comparisons between the inferred and the empirical activities are also displayed in Supplementary Fig.\ 18 for H1N1, in Supplementary Fig.\ 19 for SARS and in Supplementary Fig.\ 20 for COVID-19.

\begin{figure*}[!ht]
	\centering
	\includegraphics[width=0.8\textwidth]{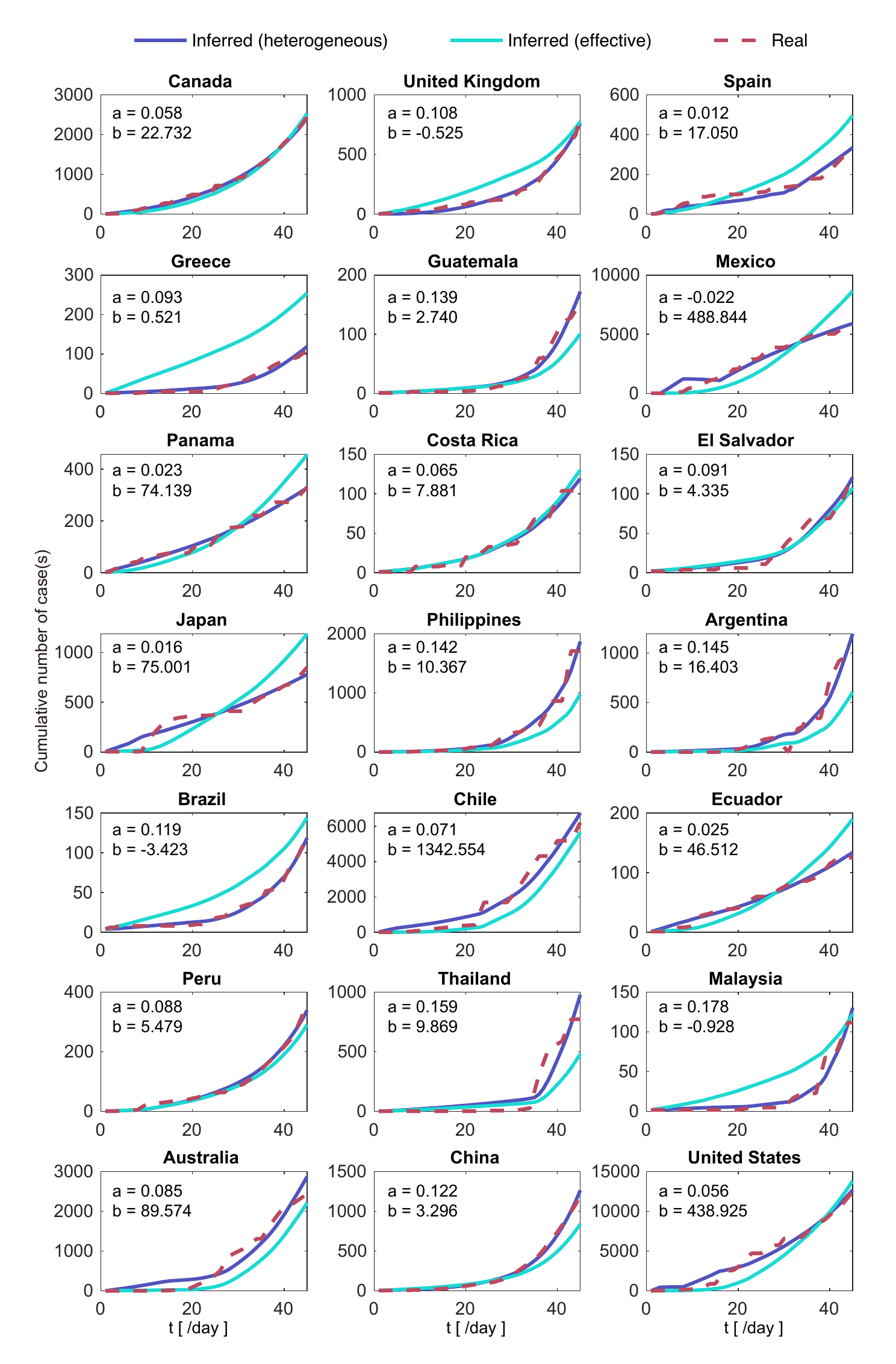}
	\caption{Comparison between empirical number of H1N1 cases, the infected numbers generated by the inferred Eq.\ (5) in the main text with identical parameters $a$ and $b$ for all nodes (\textit{i.e.}\ effective), and the infected numbers generated by the form of Eq.\ (5) yet with heterogeneous parameters $a_i$ and $b_i$ for each node $i$ (\textit{i.e.}\ heterogeneous). The parameter for effective dynamics are $a=0.074$ and $b=7.13$, and those $a_i$ and $b_i$ for heterogeneous dynamics are shown in the sub-plots respectively.}
\end{figure*}

\begin{figure*}[!ht]
	\centering
	\includegraphics[width=0.6\textwidth]{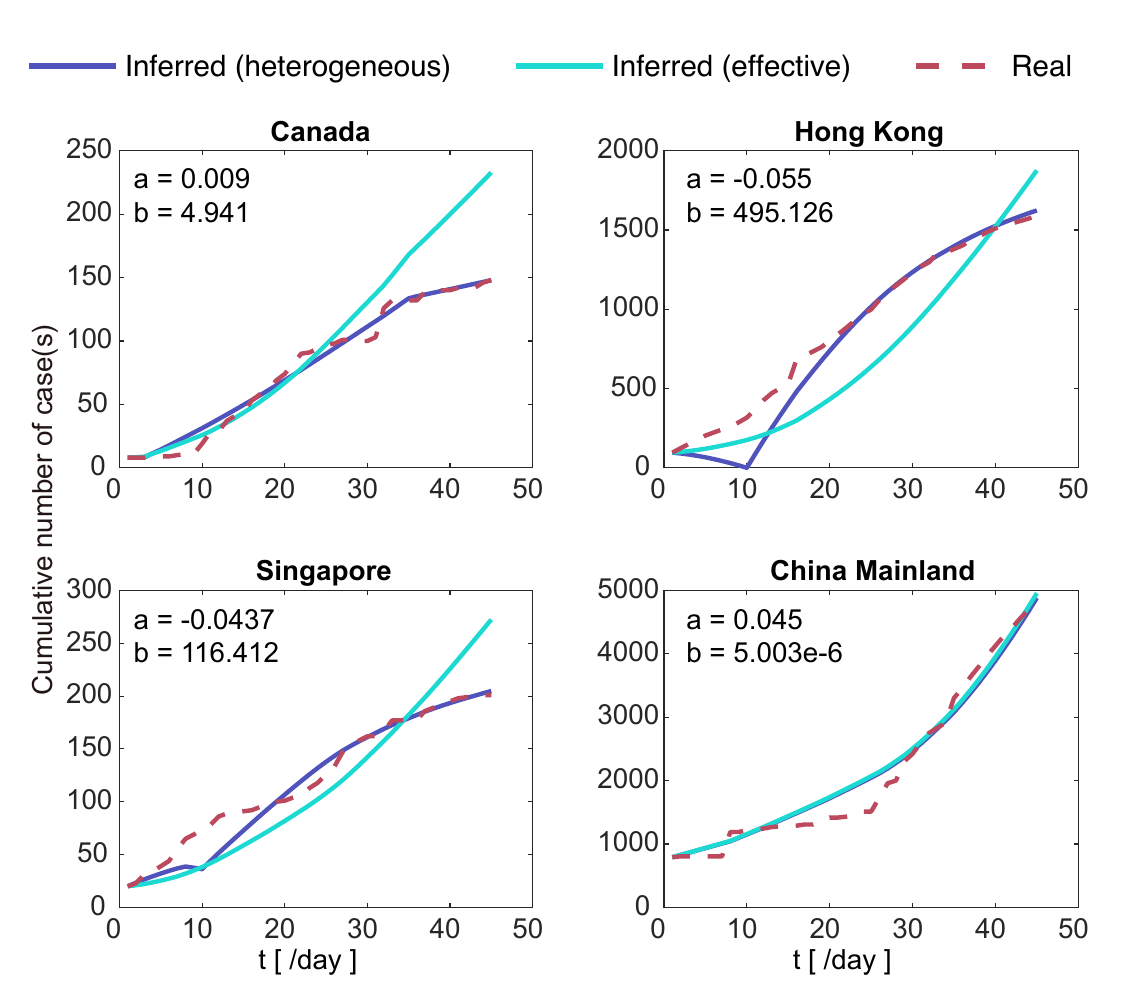}
	\caption{Comparison between empirical number of SARS cases, the infected numbers generated by the inferred Eq.\ (5) in the main text with identical parameters $a$ and $b$ for all nodes (\textit{i.e.}\ effective), and the infected numbers generated by the form of Eq.\ (5) yet with heterogeneous parameters $a_i$ and $b_i$ for each node $i$ (\textit{i.e.}\ heterogeneous). The parameter for effective dynamics are $a=0.046$ and $b=3.083$, and those $a_i$ and $b_i$ for heterogeneous dynamics are shown in the sub-plots respectively. Note that Eq.\ (5) was inferred based only on H1N1 data, hence the results in this figure demonstrate the generalizability of the inferred dynamical equation by our approach.}
\end{figure*}

\begin{figure*}[!ht]
	\centering
	\includegraphics[width=0.9\textwidth]{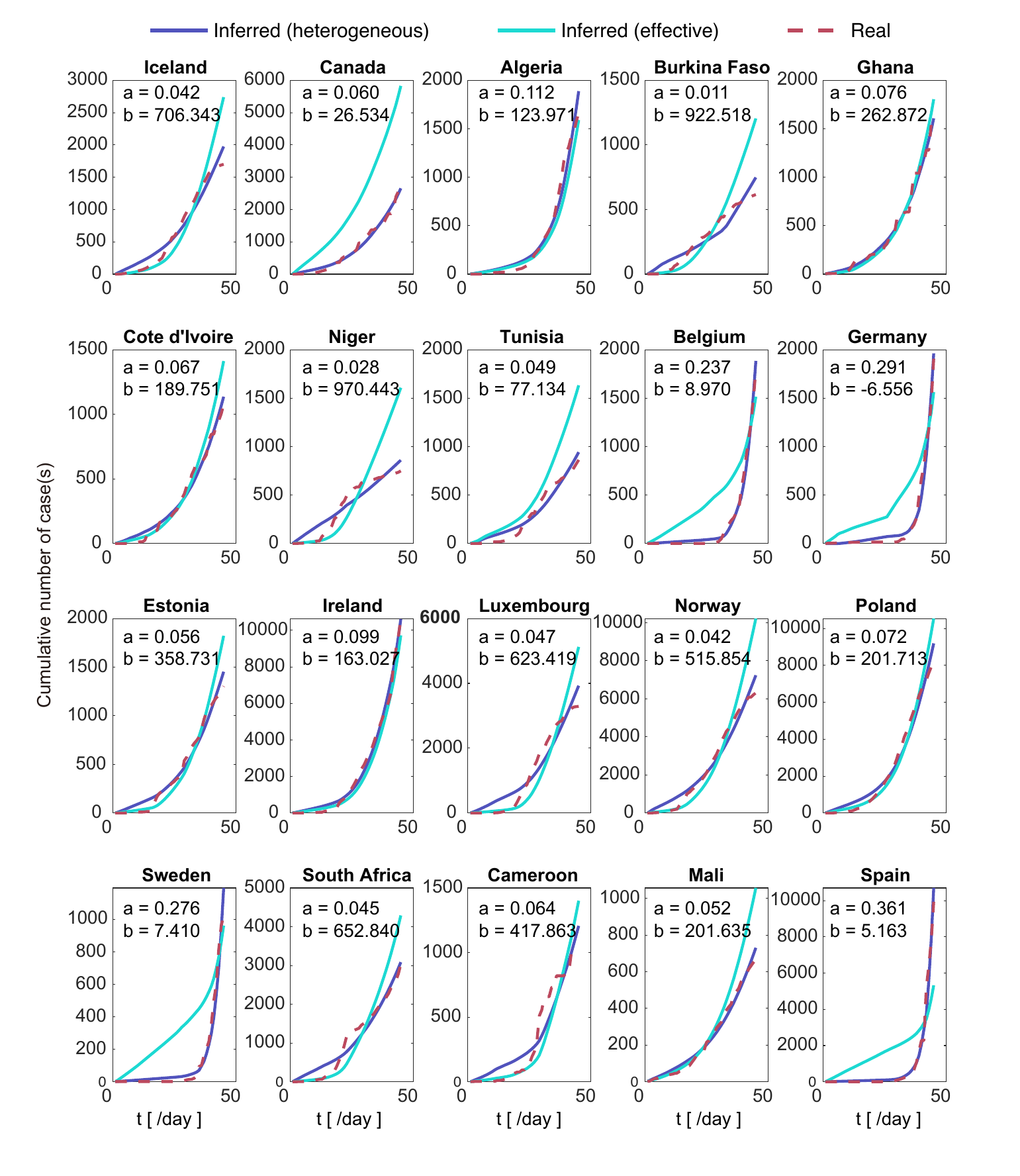}
	\caption{Comparison between empirical number of COVID-19 cases, the infected numbers generated by the inferred Eq.\ (5) in the main text with identical parameters $a$ and $b$ for all nodes (\textit{i.e.}\ effective), and the infected numbers generated by the form of Eq.\ (5) yet with heterogeneous parameters $a_i$ and $b_i$ for each node $i$ (\textit{i.e.}\ heterogeneous). The parameter for effective dynamics are $a=0.040$ and $b=105.160$, and those $a_i$ and $b_i$ for heterogeneous dynamics are shown in the sub-plots respectively. Note that Eq.\ (5) was inferred based only on H1N1 data, hence the results in this figure show that the inferred equation is also able to capture the early spreads of COVID-19.}
\end{figure*}

\end{document}